\documentclass[a4paper,fleqn]{cas-dc}

\usepackage[numbers]{natbib}
\usepackage{booktabs}
\usepackage{tabularx}
\usepackage{multirow}

\usepackage{xcolor}
\usepackage{pifont}
\newcommand{\cmark}{\textcolor{green!60!black}{\ding{51}}}  
\newcommand{\xmark}{\textcolor{red}{\ding{55}}}             
\newcommand{\pmark}{\textcolor{orange}{\ding{115}}}         

\usepackage{float}
\usepackage{subcaption}

\usepackage{algorithm}
\usepackage{algpseudocode}

\usepackage{lastpage}
\usepackage{fancyhdr}

\usepackage{tcolorbox}
\tcbuselibrary{skins,breakable}

\newtcolorbox{promptbox}[1]{
    title={#1},
    colback=gray!3,
    colframe=black,
    fonttitle=\bfseries,
    fontupper=\footnotesize,
    enhanced,
    width=\linewidth
}

\def\tsc#1{\csdef{#1}{\textsc{\lowercase{#1}}\xspace}}
\tsc{WGM}
\tsc{QE}

\begin{document}
\let\WriteBookmarks\relax
\def\floatpagepagefraction{1}
\def\textpagefraction{.001}

\shorttitle{}


\title[mode=title]{AI-Assisted Cybersecurity Policy Assessment: Evidence Grounding, Coverage Gaps, and Implications for Security Management
}

\author[1]{Bikash Saha}

\cormark[1]


\ead{bikash@cse.iitk.ac.in}

\ead[url]{}

\credit{Conceptualization of this study, Data Curation, Investigation, Methodology, Writing - Original Draft, Project administration, Writing - Review \& Editing, Visualization}

\affiliation[1]{organization={Indian Institute of Technology Kanpur},
            country={India}}

\author[2]{Sandeep Kumar Shukla}


\ead{sandeeps@iiit.ac.in}

\ead[url]{}

\credit{Resources, Validation, Writing - Review \& Editing, Supervision}

\affiliation[2]{organization={International Institute of Information Technology Hyderabad},
            country={India}}

\cortext[1]{Corresponding author}
\fntext[1]{}





\begin{abstract}
Cybersecurity policy assessment requires reviewers to determine whether organizational policies adequately address established security controls and to identify areas where further policy development is needed. This process is challenging because relevant evidence may be distributed across multiple documents, expressed using different terminology, or address only part of a control requirement. In this study, we examine how evidence-grounded automated analysis can support control-level cybersecurity policy assessment. We present PACE (Policy Assessment with Control Evidence), an approach that retrieves relevant policy evidence, distinguishes between full, partial, and absent coverage, identifies missing policy elements, and provides recommendations and evidence-linked explanations for expert review. We evaluate PACE using 1,007 NIST SP~800-53 controls across two real-world organizational policy corpora. The approach achieves F1-scores of 88.54\% and 82.31\% and consistently outperforms the evaluated baseline methods. 
Our results show that semantic evidence retrieval substantially improves policy assessment over single-stage analysis. 
Expert evaluation also indicates that the identified policy gaps and supporting evidence are generally consistent with professional assessments. These findings highlight the value of evidence traceability, explicit treatment of partial policy coverage, and human review when automated methods are used to support cybersecurity policy assessment and governance.
\end{abstract}

\maketitle

\pagestyle{fancy}
\fancyhf{}

\fancyfoot[L]{\textit{Preprint submitted to Elsevier}}
\fancyfoot[R]{Page \thepage\ of \pageref{LastPage}}

\renewcommand{\headrulewidth}{0pt}
\renewcommand{\footrulewidth}{0.4pt}

\section{Introduction}
\label{sec:introduction}

Organizations increasingly depend on information systems, cloud services,
and connected infrastructure to support their daily operations. As this
dependence grows, cybersecurity policies play an important role in defining
security requirements, assigning responsibilities, and establishing
organizational accountability~\cite{knapp2009information,cram2017organizational}.
These policies also provide a basis for risk management, internal oversight,
and security audits~\cite{slapnivcar2022effectiveness,edwards2024comprehensive}.
To determine whether their policies address expected security practices,
organizations commonly assess them against established frameworks such as
NIST SP~800-53~\cite{force2020security,force2022assessing}.

In practice, however, assessing cybersecurity policies against security
controls is difficult. Relevant information may be distributed across
multiple documents, expressed using organization-specific terminology, or
stated only partially~\cite{cram2017organizational,chithaluru2020organization}.
A policy may, for example, require periodic access reviews without identifying
who is responsible for conducting them or how frequently they should occur.
In such cases, the policy is relevant to the control, but the available
documentation may still be incomplete. Reviewers must therefore do more than
search for matching security terms. They must locate relevant evidence,
interpret it in the context of the control, determine whether the documented
coverage is sufficient, and identify what remains missing.

These judgments have practical consequences for cybersecurity governance.
Treating incomplete documentation as sufficient can create false assurance
that a security requirement has been adequately addressed. Conversely,
treating relevant but incomplete policy content as entirely absent can lead
to unnecessary review and policy-revision effort. The distinction between
complete, partial, and absent coverage is therefore important for both audit
quality and the efficient use of security resources. It is also important to
separate documented policy coverage from operational control effectiveness.
A policy may adequately describe a required security practice without
demonstrating that the practice has been implemented effectively, while a
practice may exist operationally without being documented sufficiently for
audit or governance purposes.

Cybersecurity policy assessment has traditionally relied on manual review and
professional judgment. Reviewers interpret control requirements, locate
relevant policy provisions, assess whether the available documentation is
sufficient, and record the basis for their conclusions
~\cite{force2022assessing,slapnivcar2022effectiveness}. As policy collections
grow in size and complexity, this process becomes increasingly time-consuming.
Computational methods have therefore been explored for requirement extraction,
compliance checking, semantic retrieval, and policy analysis
~\cite{breaux2008analyzing,xiao2012automated,mandal2018automating,
saha2025parag,hassani2024rethinking}. These methods can reduce the effort
needed to locate relevant information, but finding related text does not by
itself establish that a control has been fully addressed.

This limitation is particularly important in cybersecurity policy review.
A retrieved passage may discuss the correct security topic while omitting
important elements such as scope, responsibility, procedures, review
frequency, enforcement, or exception handling. An assessment method must
therefore consider not only whether relevant evidence exists, but also
whether that evidence is sufficient for the control being examined.
When coverage is incomplete, reviewers also need to understand which policy
elements are missing and what type of clarification or revision may be
required. Existing approaches provide useful support for policy information
extraction, compliance reasoning, semantic retrieval, and policy-as-code
generation~\cite{rodriguez2024large,hassani2024rethinking,saha2025parag,
madan2025argen,romeo2025arpaccino}, but these capabilities are often examined
as separate tasks. Less attention has been given to the complete
control-level assessment process in which evidence must be located,
evaluated for completeness, linked to identified deficiencies, and presented
in a form that can be checked by a security professional.

The ability to verify an automated assessment is especially important in
security governance. Responsibility for an audit finding or policy decision
ultimately remains with the reviewer rather than with the assessment tool.
The supporting evidence should therefore remain visible, and the reviewer
should be able to determine why a particular coverage judgment was suggested.
Prior research on human--AI decision making similarly emphasizes that
automated support is most useful when it complements professional judgment
rather than replacing it~\cite{jarrahi2018humanai,shrestha2019organizational}.
Research on explainability also highlights the importance of presenting
information that allows users to understand and evaluate the basis of an
automated recommendation~\cite{coussement2024explainable}. In the context of
cybersecurity policy assessment, this means linking each suggested assessment
to the policy evidence on which it is based and making incomplete coverage
explicit rather than reducing the outcome to a binary compliance judgment.

Against this background, we investigate how evidence-grounded automated
analysis can support cybersecurity policy assessment while preserving the
role of professional review. We present PACE (Policy Assessment with Control Evidence), a control-level policy
assessment approach that retrieves relevant evidence from an organizational
policy corpus and evaluates the extent to which that evidence addresses a
security control. PACE distinguishes among
\texttt{FULLY\_COVERED}, \texttt{PARTIALLY\_COVERED}, and
\texttt{NOT\_COVERED} cases. For controls with incomplete or absent
documentation, it identifies missing policy elements and provides
policy-level recommendations. Each assessment retains the corresponding
policy evidence so that reviewers can inspect the basis of the suggested
coverage state. PACE is intended to support policy review rather than
make an autonomous compliance determination.

We study this approach using 1,007 controls from NIST SP~800-53 and two
independent real-world organizational policy corpora. Expert reference
assessments are constructed for both organizations, allowing us to compare
automated assessments with human judgments at the control level. The study
examines four questions. First, how accurately can evidence-grounded analysis
distinguish full, partial, and absent policy coverage? Second, how much does
the quality of retrieved evidence affect the resulting assessment? Third,
does separating evidence retrieval, coverage assessment, and gap
identification improve reliability compared with less structured approaches?
Finally, how useful are the resulting gap diagnoses, recommendations, and
evidence-linked explanations for expert review? We also examine performance
across different organizational policy environments and model backbones to
understand how sensitive the results are to these factors.

Our evaluation shows that PACE achieves F1-scores of 88.54\% and
82.31\% across the two organizational policy corpora and performs better than
the evaluated single-stage, lexical, dense-retrieval, and document-level
baselines. The results also show that semantic retrieval provides a
substantial improvement over approaches that rely on lexical matching or
unstructured document-level analysis. Importantly, the findings indicate
that retrieval alone is not sufficient. Explicitly evaluating whether
retrieved evidence is complete or only partial further improves the quality
of control-level assessments. Expert review of the diagnostic outputs also
shows that the identified gaps and supporting evidence generally align with
professional assessment.

These results have implications beyond model performance. They show that
automated cybersecurity policy assessment is most useful when it preserves
the connection between a control, the supporting policy evidence, and the
resulting assessment. They also highlight the importance of representing
partial coverage explicitly. From a security-management perspective, this
distinction can help reduce the risk of false assurance while avoiding the
loss of useful evidence that would result from treating every incomplete
control as entirely uncovered. At the same time, differences across
organizations and model backbones reinforce the need for local validation
and continued human review before such assessments are used to support
audit preparation or policy revision.
This paper makes the following contributions:

\begin{itemize}

    \item We provide an empirical study of automated cybersecurity policy
    assessment using 1,007 NIST SP~800-53 controls and two real-world
    organizational policy corpora, with expert reference assessments for
    both organizations.

    \item We examine how evidence retrieval and explicit treatment of
    partial coverage affect the reliability of control-level policy
    assessment, including comparisons with single-stage, lexical,
    dense-retrieval, and document-level approaches.

    \item We evaluate policy-gap identification, recommendation usefulness,
    and evidence-linked explanations as practical forms of support for
    security professionals reviewing incomplete or absent control coverage.

    \item We identify implications for cybersecurity policy review and
    governance, particularly the importance of evidence traceability,
    distinguishing partial from complete coverage, organization-specific
    validation, and retaining human oversight of the final assessment.

\end{itemize}

\section{Related Work}
\label{sec:pro_related_work}

Research on cybersecurity policy assessment draws from several related
areas, including security governance and compliance, automated analysis
of policy documents, and the use of AI to support professional review.
This section first discusses cybersecurity policy and compliance
assessment, followed by recent approaches for automated and
retrieval-based policy analysis. We then consider the role of evidence,
explainability, and human oversight in security assessment before
identifying the gap addressed in this work.

\subsection{Cybersecurity Policy, Governance, and Compliance}
\label{sec:rw_compliance}

Cybersecurity policies establish how organizations expect security
responsibilities, procedures, and controls to be managed. They provide
a formal basis for communicating security expectations, assigning
responsibilities, supporting risk management, and demonstrating
alignment with external requirements. Assessing these policies against
standards and regulatory requirements is therefore an important part of
security governance and audit practice.

Policy assessment is difficult because the language used in security
standards does not always correspond directly to the language used in
organizational policies. Requirements may also be distributed across
several documents or expressed at different levels of detail. A control
that appears to be addressed in one section may depend on responsibilities,
procedures, or review requirements defined elsewhere. This makes policy
assessment a task of interpretation as well as information retrieval.

Earlier computational approaches sought to make compliance analysis more
systematic through rules, ontologies, keyword matching, and structured
representations. Breaux and Antón~\cite{breaux2008analyzing}, for
example, showed how regulatory requirements could be extracted and
represented for further analysis. Such approaches provide useful
structure when requirements can be expressed through predefined rules
or relationships. Their effectiveness can be more limited, however,
when equivalent requirements are described using different terminology
or when policy statements are incomplete or distributed across several
sources.

More recent research has considered compliance within the broader
context of organizational governance. Regulatory technologies now
support activities such as compliance monitoring, risk assessment,
reporting, and institutional oversight
~\cite{micklitz2024compliance,cave2026regulating}. Empirical research
also shows that changes in the external regulatory environment can
affect organizational security behavior. Wang et
al.~\cite{wang2024alert}, for example, report changes in IT and
cybersecurity investment following the enactment of state
anti-phishing legislation, with organizational responses varying
according to industry risk and IT capability. Such findings illustrate
the connection between external security requirements and internal
governance practices.

The growing use of computational methods in this area has also raised
questions about adaptability, trust, and evaluation. Jain et
al.~\cite{jain2025complexity} review the use of AI and natural language
processing for compliance-related tasks and identify several challenges
that remain when these methods are applied in organizational settings.
Other work examines how regulations, standards, certification
mechanisms, and governance principles are translated into organizational
practice~\cite{agarwal2025five,ulnicane2021framing,mueller2025s}.
Together, this literature shows that compliance assessment is not
limited to identifying whether a security topic is mentioned. It also
requires an understanding of how requirements are represented,
documented, and governed within the organization.

Another line of research focuses on converting governance requirements
into machine-readable or executable forms. ArGen
~\cite{madan2025argen} uses policy-as-code mechanisms to align system
behavior with governance requirements. Policy-as-Prompt
~\cite{kholkar2025policy} translates governance policies into runtime
guardrails for AI agents, while ARPACCINO
~\cite{romeo2025arpaccino} combines retrieval and tool-supported
reasoning for policy-as-code generation and validation. Tummala et
al.~\cite{tummala2026autonomous} use logic-based representations to
reason about policies, compliance states, and associated penalties.

These approaches are useful for converting requirements into rules,
supporting enforcement, or reasoning about policy-aware system
behavior. The problem considered in this work is different. We focus on
the assessment of existing human-written cybersecurity policies and ask
whether those policies provide sufficient documented evidence for a
given security control.

\subsection{Automated and AI-Assisted Cybersecurity Policy Analysis}
\label{sec:rw_ai_compliance}

Advances in natural language processing have made it possible to analyze
policy and compliance documents with less dependence on fixed keywords
or manually defined rules. Large language models are increasingly used
to extract information from policies, interpret requirements, and
support compliance-related analysis. Rodriguez et
al.~\cite{rodriguez2024large}, for example, use large language models to
extract privacy practices from policy documents. Hassani et
al.~\cite{hassani2024rethinking} examine the use of language-model
reasoning for compliance automation and argue for more structured forms
of analysis than isolated classification tasks.

Retrieval-augmented approaches extend this idea by first identifying
relevant source material and then using the retrieved content during
analysis. This is particularly useful for organizational policies
because the relevant evidence may be distributed across long and
heterogeneous document collections. PARAG~\cite{saha2025parag}, for
example, uses semantically indexed security-policy content to answer
questions about organizational policies. Retrieving policy evidence
before generating an answer can reduce dependence on information stored
in the model itself and makes it possible to connect an output to the
underlying organizational documents.

Retrieval alone, however, does not resolve the main difficulty of
policy assessment. A passage may be relevant to a security control
without providing enough information to show that the control is fully
addressed. A policy may state that access reviews must occur, for
example, but leave the responsible role, review frequency, or scope
undefined. Similarly, a policy may describe a security objective in
general terms without specifying the procedures or accountability
mechanisms required by the corresponding control.

For this reason, cybersecurity policy assessment requires a distinction
between \emph{relevance} and \emph{sufficiency}. The first concerns
whether a policy passage relates to the control under review. The second
concerns whether the available policy evidence addresses the important
elements of that control. This distinction becomes particularly
important when the available documentation is incomplete. Treating
every relevant passage as evidence of complete coverage may lead to
false assurance, whereas treating incomplete evidence as entirely
absent can overlook policy provisions that are already in place.

A useful assessment process should therefore do more than retrieve
policy text or produce a final coverage label. It should preserve the
relevant evidence, distinguish complete from partial coverage, identify
the elements that remain undocumented, and provide enough context for a
reviewer to examine the result.

Table~\ref{tab:rw_comparison} summarizes representative approaches
according to capabilities that are relevant to cybersecurity policy
assessment. The comparison considers whether an approach provides
document-level evidence, distinguishes intermediate or incomplete
assessment states, identifies deficiencies, provides corrective
guidance, supports human review, and preserves traceability to the
underlying policy material.

\begin{table*}[htbp]
\centering
\caption{Capabilities of representative automated policy and compliance
approaches relevant to cybersecurity policy assessment.}
\label{tab:rw_comparison}
\scriptsize
\setlength{\tabcolsep}{4pt}
\begin{tabularx}{\textwidth}{p{2.3cm} X c c c c c c}
\toprule
\textbf{Work}
& \textbf{Primary task}
& \shortstack{\textbf{Policy}\\\textbf{evidence}}
& \shortstack{\textbf{Partial}\\\textbf{coverage}}
& \shortstack{\textbf{Gap}\\\textbf{identification}}
& \shortstack{\textbf{Corrective}\\\textbf{guidance}}
& \shortstack{\textbf{Human}\\\textbf{review}}
& \textbf{Traceability} \\
\midrule

\cite{rodriguez2024large}
& Policy information extraction
& \xmark & \xmark & \xmark & \xmark & \xmark & \xmark \\

\cite{hassani2024rethinking}
& Compliance reasoning
& \pmark & \pmark & \pmark & \pmark & \pmark & \pmark \\

\cite{saha2025parag}
& Policy question answering
& \cmark & \xmark & \xmark & \xmark & \xmark & \pmark \\

\cite{madan2025argen}
& Policy rule generation and enforcement
& \pmark & \xmark & \xmark & \pmark & \pmark & \cmark \\

\cite{kholkar2025policy}
& Runtime policy governance
& \pmark & \xmark & \xmark & \xmark & \pmark & \cmark \\

\cite{romeo2025arpaccino}
& Policy-as-code generation and validation
& \cmark & \pmark & \pmark & \pmark & \pmark & \cmark \\

\cite{tummala2026autonomous}
& Policy-aware compliance reasoning
& \xmark & \cmark & \pmark & \xmark & \pmark & \cmark \\

\midrule

\textbf{PACE}
& Control-level policy assessment
& \cmark & \cmark & \cmark & \cmark & \cmark & \cmark \\

\bottomrule
\multicolumn{8}{l}{
\footnotesize
\cmark\ = supported, \pmark\ = partial or indirect support,
\xmark\ = not a primary feature.
}
\end{tabularx}
\end{table*}

\subsection{Evidence, Explainability, and Human Oversight}
\label{sec:rw_human_review}

Automated policy assessment is most useful when security professionals
can examine the basis of the resulting judgment. This is especially
important in cybersecurity governance, where an incorrect assessment
can affect audit preparation, policy revision, and decisions about where
security resources should be directed. A coverage label without access
to the underlying policy evidence provides limited support for verifying
whether the assessment is reasonable.

Research on human--AI decision making provides a useful perspective on
this issue. Jarrahi~\cite{jarrahi2018humanai} describes human and
machine capabilities as complementary, particularly in situations that
combine large amounts of information with contextual judgment.
Shrestha et al.~\cite{shrestha2019organizational} similarly examine
different ways in which human and algorithmic decisions can be combined
within organizational processes. Their work distinguishes settings in
which decisions are delegated to automated systems from those in which
automation provides information or recommendations while a human retains
responsibility for the final judgment.

This distinction is relevant to cybersecurity policy assessment.
Automated analysis can help locate and organize information across a
large collection of policy documents, but deciding whether the available
evidence is sufficient may still require knowledge of organizational
context and the intent of the security requirement. The role of
automation is therefore better understood as assisting the reviewer
rather than replacing professional judgment.

Explainability is closely related to this requirement.
Coussement et al.~\cite{coussement2024explainable} highlight the role
of explanations in helping users make informed decisions with AI-based
systems. In security-related settings, transparency is particularly
important because reviewers may need to justify an assessment to other
stakeholders or revisit it during an audit. Work on explainable AI in
cyber-resilience similarly emphasizes the relationship between
explanation and transparency in organizational security processes
~\cite{sadeghi2024explainable}.

For policy assessment, explanation should begin with the evidence
itself. The reviewer should be able to see which policy statements
contributed to a suggested assessment and which required elements were
judged to be missing. Maintaining this connection between the source
document and the resulting judgment makes it possible to verify,
challenge, or revise the automated output. It also reduces the risk
that general model knowledge is mistaken for evidence of what an
organization has actually documented.

Human oversight is therefore not an additional feature applied after
the assessment has been produced. It is part of the assessment process.
Useful automated support should organize the available evidence and
make deficiencies easier to identify while leaving the final
interpretation and governance decision with the security professional.

\subsection{Research Gap}
\label{sec:rw_gap}

Existing research provides several of the capabilities needed for
automated cybersecurity policy assessment. Earlier compliance methods
provide structured representations of requirements, while more recent
approaches support semantic retrieval, policy reasoning, information
extraction, and policy-as-code generation. Retrieval-augmented methods
also make it easier to ground automated analysis in organizational
documents rather than relying solely on a model's general knowledge.

What remains less well studied is how these capabilities perform when
the objective is to assess the \emph{completeness of documented policy
coverage} at the level of individual security controls. Retrieving a
relevant policy passage is only the first step. Reviewers must still
determine whether the evidence fully addresses the control, addresses
only part of it, or provides no meaningful coverage. When documentation
is incomplete, they also need to know which elements are missing and
whether the resulting recommendation is consistent with both the
control and the available policy evidence.

A second gap concerns the way automated assessments are presented for
professional review. Much of the existing literature evaluates
information extraction, retrieval, or compliance reasoning as separate
tasks. Less attention has been given to maintaining a traceable
connection among the security control, the retrieved evidence, the
coverage assessment, and the identified policy gaps. This connection is
important because cybersecurity policy assessment can influence audit
findings and policy changes, and reviewers must be able to verify the
basis of an automated result.

This study addresses these gaps by examining evidence-grounded,
control-level cybersecurity policy assessment across two real-world
organizational policy corpora. We evaluate whether relevant policy
evidence can be retrieved reliably, whether full, partial, and absent
coverage can be distinguished, and whether incomplete assessments can
be accompanied by meaningful gap diagnoses and policy-level
recommendations. We also examine whether these outputs remain
traceable to the underlying policy evidence and suitable for expert
review. In this way, the study focuses not only on whether automated
methods can produce an assessment, but also on whether that assessment
provides the information needed for responsible use in cybersecurity
policy review.

\section{Cybersecurity Policy Assessment Framework}
\label{sec:assessment_framework}

\subsection{Assessment Scope}

This study focuses on whether an organization's written cybersecurity
policies provide sufficient documented evidence for the requirements
specified by a security control. The assessment concerns policy
documentation rather than the operational effectiveness of the control.
A control may therefore be adequately described in policy without being
implemented effectively in practice, and operational practices may exist
without being sufficiently documented.

The primary unit of analysis is an individual security control together
with the organizational policy evidence relevant to that control. For
each control, the assessment asks whether the available documentation
provides sufficient evidence of the required security practice.

\subsection{Policy Coverage Categories}

We distinguish three levels of documented policy coverage:

\begin{itemize}

    \item \texttt{FULLY\_COVERED}: The available policy evidence
    addresses the control objective and provides the main information
    required to document the corresponding security practice.

    \item \texttt{PARTIALLY\_COVERED}: Relevant policy evidence exists,
    but one or more important elements of the control are incomplete,
    unclear, or absent.

    \item \texttt{NOT\_COVERED}: The policy corpus does not provide
    sufficient evidence addressing the control objective.

\end{itemize}

The distinction between full and partial coverage is important because
the presence of relevant policy language does not necessarily demonstrate
that a control has been adequately documented. For example, a policy may
require periodic access reviews while leaving the responsible role or
review frequency unspecified. Such a policy contains relevant evidence,
but important elements required for complete coverage remain absent.

\subsection{Challenges in Control-Level Policy Assessment}

Three characteristics of organizational policy collections make this
assessment difficult. First, evidence relevant to a single control may
be distributed across multiple documents or sections. Responsibilities
may be defined in one policy while procedures or review requirements
appear elsewhere.

Second, security controls and organizational policies often use
different terminology. Standardized controls describe requirements in
general terms, whereas organizations express similar practices using
their own language and governance structures. Keyword-based matching
may therefore miss relevant evidence.

Third, relevant policy statements may provide only partial coverage.
A policy can address the general intent of a control while omitting
important details such as responsibility, scope, procedures, review
frequency, enforcement, or exception handling. Assessment therefore
requires both the identification of relevant evidence and an evaluation
of whether that evidence is sufficient.

These characteristics motivate an assessment process that preserves the
source evidence, distinguishes partial from complete coverage, and
identifies missing policy elements when documentation is incomplete.
The following section describes how this assessment is implemented and
evaluated in our study.

\section{Methodology}
\label{sec:pro_methodology}

This section describes the procedure used to assess documented
cybersecurity policy coverage. PACE operates at the level of an
individual security control. For each control, it first identifies
relevant evidence from the organizational policy collection and then
examines whether that evidence provides full, partial, or no documented
coverage. When the available evidence is incomplete, the procedure
identifies the missing policy elements and provides policy-level
recommendations. The control, supporting evidence, coverage assessment,
identified gaps, and recommendations are retained together so that the
result can be reviewed by a security professional.

\begin{figure*}[t]
\centering
\includegraphics[width=0.8\linewidth]{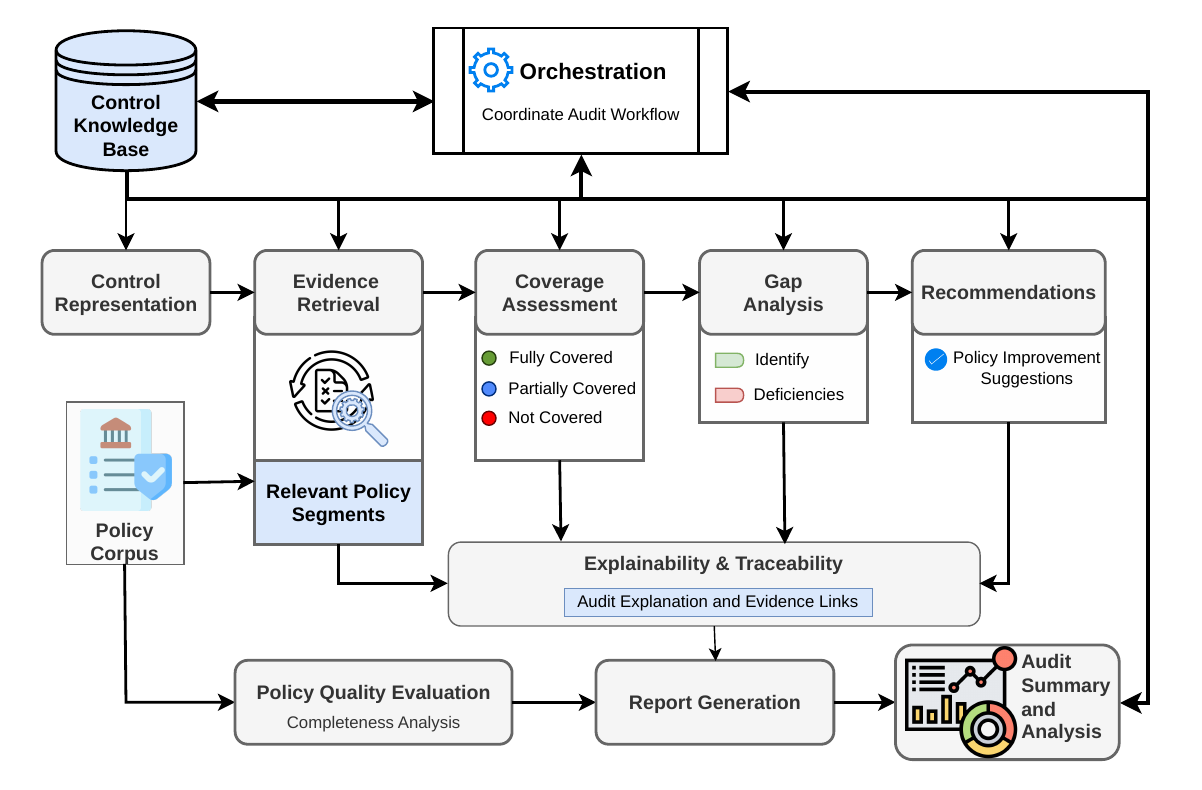}
\caption{Overview of the PACE cybersecurity policy assessment
procedure. For each security control, relevant organizational policy
evidence is retrieved and assessed for full, partial, or absent
documented coverage. Controls with incomplete or absent coverage are
further examined to identify missing policy elements and possible
policy-level improvements. The evidence and resulting assessment are
retained for expert review.}
\label{fig:method}
\end{figure*}

Figure~\ref{fig:method} summarizes the assessment procedure. The process
is intentionally evidence-centered. Each suggested assessment remains
linked to the organizational policy text on which it is based. This is
important because the objective is to assess what the organization has
documented rather than infer practices that are not supported by the
policy corpus.

\subsection{Policy Corpus and Control Preparation}
\label{subsec:policy_preprocessing}

The assessment begins with two sources of information: the
organization's cybersecurity policy collection and the security controls
against which those policies are evaluated.

Organizational policy documents are first converted into plain text.
Repeated headers, footers, page numbers, and formatting artifacts are
removed, while structural information such as section headings, lists,
tables, and appendices is retained where possible. Preserving this
structure is useful because the meaning of a policy statement may depend
on the section or governance context in which it appears.

Following the document-processing configuration used in earlier
policy-analysis work~\cite{saha2025parag}, each cleaned document is
divided into segments of 512 tokens with an overlap of 50 tokens.
The overlap reduces the chance that related policy statements are
separated at segment boundaries. Each segment is stored together with
its source information, including the document name, section heading,
page or source location, and a unique segment identifier. These
metadata are retained throughout the assessment so that retrieved
evidence can later be traced back to the original policy document.

The policy segments are represented using the
\texttt{all-MiniLM-L6-v2} sentence-embedding model
~\cite{reimers2019sentencebert} and indexed for semantic retrieval.

The security-control collection is constructed from NIST SP~800-53.
For each control, we retain its identifier, title, control family, and
control statement. The original control statement is used throughout
the assessment as the reference requirement. Using the same control
text for retrieval and coverage assessment provides a consistent basis
for comparing results across organizations and experimental settings.

\subsection{Policy Evidence Retrieval}
\label{subsec:evidence_retrieval}

For each security control, PACE retrieves policy passages that are
most relevant to the control requirement. The purpose of this step is
not to determine whether the control is satisfied, but to identify the
organizational documentation that should be considered when making that
assessment.

The control information is used as the retrieval query, and the most
relevant policy segments are selected from the indexed policy corpus.
Each retrieved segment retains its document and source metadata. As a
result, reviewers can inspect not only the policy text selected by the
method but also where that text appears in the original organizational
documents.

Semantic retrieval is used because security controls and organizational
policies do not always use the same terminology. A standardized control
may describe a requirement using general security language, while an
organization may express the same practice using local roles, procedures,
or governance terminology. Semantic matching therefore provides a way
to identify relevant passages even when there is limited lexical overlap
between the control and the policy text.

\subsection{Assessment of Documented Policy Coverage}
\label{subsec:coverage_assessment}

The retrieved evidence is then examined against the corresponding
security-control statement. The objective is to determine how completely
the organization's written policies address the control requirement.

The assessment uses the three coverage categories introduced in
Section~\ref{sec:assessment_framework}:
\texttt{FULLY\_COVERED}, \texttt{PARTIALLY\_COVERED}, and
\texttt{NOT\_COVERED}. A control is considered fully covered when the
available policy evidence addresses the control objective and provides
the main information needed to document the corresponding security
practice. Partial coverage is assigned when relevant evidence exists but
important elements remain unclear or absent. A control is considered
not covered when the policy corpus does not provide meaningful evidence
for the control objective.

The assessment considers the control requirement together with all
retrieved policy evidence rather than relying on the presence of an
individual matching passage. This is necessary because the information
needed for a single control may be distributed across multiple policy
sections. For example, one document may define the required security
practice while another assigns responsibility for carrying it out.

Particular attention is given to the distinction between full and
partial coverage. A policy statement may address the general subject of
a control but omit details that are important for documented governance,
such as responsibility, scope, procedures, review frequency, enforcement,
or exception handling. Such cases are treated as partially covered
rather than fully covered.

These labels refer only to documented policy coverage. They do not
indicate whether the corresponding security control has been implemented
effectively in practice. PACE therefore evaluates the completeness
of organizational policy documentation rather than operational
compliance or technical control effectiveness.

\subsection{Gap Identification and Policy Recommendations}
\label{subsec:gap_recommendation}

When a control is assessed as \texttt{PARTIALLY\_COVERED} or
\texttt{NOT\_COVERED}, the available evidence is examined to identify
the elements that prevent complete documented coverage. The purpose of
this step is to make an incomplete assessment more informative than a
coverage label alone.

The identified gaps depend on the requirements of the individual control
and the evidence available in the policy corpus. Common examples include
an unassigned responsibility, incomplete procedures, unclear scope,
missing review requirements, weak accountability or enforcement
language, undefined exception handling, and missing documentation or
reporting obligations.

Gap identification is tied directly to the relationship between the
security-control requirement and the retrieved policy evidence. It is
not intended to produce a general list of cybersecurity best practices.
For example, if an organizational policy requires privileged accounts
to be reviewed but does not specify who performs the review or how often
it takes place, the resulting assessment may identify responsibility
and review frequency as missing elements.

For each identified gap, PACE also produces a policy-level
recommendation describing how the documentation could be strengthened.
Recommendations focus on what should be added, clarified, assigned, or
formalized in the policy. They remain at the governance level and do not
recommend particular security products or technical configurations
unless such detail is explicitly required by the corresponding control.

This distinction is important because the objective of the study is to
evaluate policy documentation rather than prescribe implementation
choices. A recommendation may therefore suggest that an organization
define responsibility for periodic access reviews and state the review
frequency, but it would not normally recommend a particular identity
management platform for conducting those reviews.

\subsection{Staged Assessment Procedure}
\label{subsec:assessment_procedure}

The assessment is performed in a sequence of stages rather than asking
for the control label, identified gaps, recommendations, and explanation
in a single step. Evidence is retrieved first, followed by the coverage
assessment. Further analysis depends on the resulting coverage state.

Controls assessed as \texttt{FULLY\_COVERED} do not proceed to gap
identification or recommendation generation because no missing policy
element has been identified. Controls assessed as
\texttt{PARTIALLY\_COVERED} or \texttt{NOT\_COVERED} undergo additional
analysis to determine what is missing and how the documented coverage
could be improved.

This staged procedure also makes it possible to check whether outputs
remain consistent with one another. A reported policy gap should be
supported by the difference between the control requirement and the
available evidence, while a recommendation should correspond to an
identified gap. These relationships help limit recommendations that are
not supported by the underlying policy assessment.

Separating these stages also allows us to examine experimentally whether
structured assessment provides an advantage over approaches that
retrieve evidence and produce the complete assessment in a single step.

\subsection{Evidence-Linked Assessment Record}
\label{subsec:assessment_record_report}

For each security control, the assessment produces a structured record
that brings together the information needed for expert review.
Table~\ref{tab:assessment_record} summarizes the information retained for
each control.

The record includes the original control statement, the retrieved policy
evidence and its source information, the suggested coverage category,
and any identified gaps or policy-level recommendations. It also
contains a concise explanation connecting the coverage assessment to
the available evidence.

The purpose of the explanation is not to provide an independent
justification for an automated decision. Instead, it helps the reviewer
understand which evidence contributed to the suggested assessment and
which missing elements prevented a stronger coverage judgment. The
reviewer can therefore compare the assessment directly with the cited
policy material and accept, revise, or reject the suggested result.

\begin{table}[t]
\centering
\caption{Information retained for each control-level cybersecurity
policy assessment.}
\label{tab:assessment_record}
\begin{tabular}{p{0.25\linewidth} p{0.65\linewidth}}
\toprule
\textbf{Field} & \textbf{Description} \\
\midrule

Control ID &
Identifier of the security control under assessment. \\

Control text &
Original security-control statement used as the reference requirement. \\

Retrieved evidence &
Relevant policy excerpts together with document and source information. \\

Coverage assessment &
One of \texttt{FULLY\_COVERED},
\texttt{PARTIALLY\_COVERED}, or
\texttt{NOT\_COVERED}. \\

Identified gaps &
Policy elements that are missing, unclear, or insufficiently documented. \\

Recommendation &
Policy-level guidance for addressing an identified documentation gap,
when applicable. \\

Explanation &
Evidence-linked summary connecting the control requirement, policy
evidence, coverage assessment, and identified gaps. \\

\bottomrule
\end{tabular}
\end{table}

Maintaining these elements together provides traceability from the
assessment back to the organizational policy. This is particularly
important when the results are used during policy review or audit
preparation, because a reviewer can examine the source material instead
of relying on the coverage label alone.

At the organizational level, the control-level records can also be
summarized by control family or policy area. Such aggregation can reveal
recurring documentation issues, including frequently missing ownership
assignments, review requirements, or accountability provisions. These
summaries are used to describe broader patterns in the policy corpus,
while the underlying control-level evidence remains available for
inspection.

\section{Study Design and Evaluation}
\label{sec:pro_experiments}

We evaluate PACE using two independent organizational policy
collections and expert assessments of documented coverage for NIST
SP~800-53 controls. The evaluation is designed to examine more than
overall classification performance. In addition to asking whether the
approach can distinguish full, partial, and absent policy coverage, we
examine the quality of the supporting evidence, the types of assessment
errors that occur, the usefulness of identified policy gaps and
recommendations, and the consistency of the results across different
organizational settings.

We also compare PACE with less structured assessment approaches to
understand which parts of the procedure contribute most to reliable
policy assessment. A separate robustness analysis examines how the
results change across different language-model backbones. The model
comparison is used as a sensitivity analysis rather than as the main
focus of the study.

\subsection{Study Data}
\label{sec:dataset_setup}

\paragraph{Security Controls.}

The study uses $1{,}007$ controls from NIST SP~800-53
~\cite{force2020security}, covering $20$ control families. These controls
span a broad range of cybersecurity concerns, including access control,
audit and accountability, incident response, risk assessment,
configuration management, system protection, personnel security, and
supply-chain risk management.

Each control is assessed independently against the policy documentation
of each organization. The original NIST control statement provides the
reference requirement throughout the study. The assessment therefore
asks whether the organization's written policies contain sufficient
evidence to document the security practice described by that control.

\begin{figure}[]
    \centering
    \includegraphics[width=\linewidth]{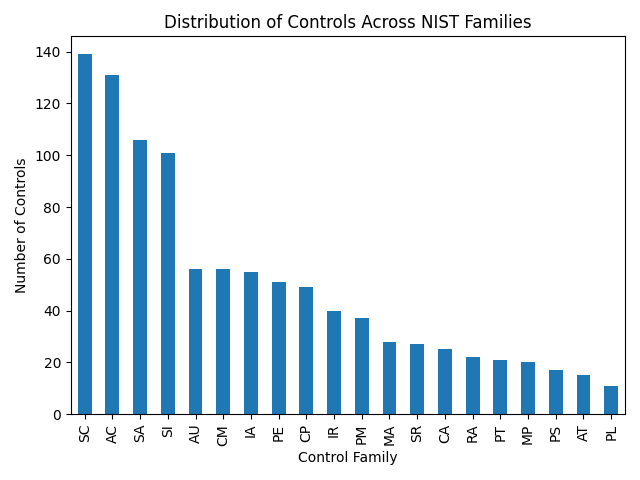}
    \caption{Distribution of evaluated controls across NIST SP~800-53
    control families. Abbreviations: AC (Access Control), AU (Audit and
    Accountability), CM (Configuration Management), CP (Contingency
    Planning), IA (Identification and Authentication), IR (Incident
    Response), MA (Maintenance), MP (Media Protection), PE (Physical and
    Environmental Protection), PL (Planning), PM (Program Management),
    PS (Personnel Security), RA (Risk Assessment), SA (System and Services
    Acquisition), SC (System and Communications Protection), SI (System
    and Information Integrity), SR (Supply Chain Risk Management), CA
    (Assessment, Authorization, and Monitoring), AT (Awareness and
    Training), and PT (Personally Identifiable Information Processing and
    Transparency).}
    \label{fig:family_distribution}
\end{figure}

Figure~\ref{fig:family_distribution} shows the distribution of the
evaluated controls across NIST SP~800-53 control families.

\paragraph{Organizational Policy Corpora.}

We evaluate the assessment procedure on two independent real-world
organizational policy corpora, referred to as OrgA and OrgB.
Organizational cybersecurity policies often contain sensitive
governance, operational, and risk-related information, which makes such
documents difficult to obtain for research purposes. Both collections
were therefore anonymized, and identifying information was removed
before analysis.

OrgA contains 24 security-policy documents comprising 356 pages and
80,123 words. OrgB contains 31 documents comprising 395 pages and
80,666 words. Table~\ref{tab:policy_stats} summarizes the main
characteristics of the two collections.

Although the corpora are similar in overall size, they differ in how
their policy content is organized and expressed. The average document
in OrgA is longer than in OrgB, reflecting differences in policy
structure and document granularity. These differences provide an
opportunity to examine whether the assessment procedure remains reliable
across organizations with different writing styles and governance
structures.

Relevant evidence is not always located in a single section or document.
A control may depend on statements appearing in several policy
documents, appendices, or cross-referenced sections. This makes the two
collections suitable for studying control-level assessment under
realistic documentation conditions.

\begin{table}[htbp]
    \centering
    \begin{tabular}{lrr}
    \toprule
    \textbf{Metric} & \textbf{OrgA} & \textbf{OrgB} \\
    \midrule
    Number of Documents & 24 & 31 \\
    Total Pages & 356 & 395 \\
    Total Words & 80,123 & 80,666 \\
    Average Words per Document & 3338.46 & 2602.13 \\
    \bottomrule
    \end{tabular}
    \caption{Structural characteristics of the two organizational policy
    corpora.}
    \label{tab:policy_stats}
\end{table}

\subsection{Expert Reference Assessments}
\label{sec:ground_truth}

To evaluate the control-level assessments, we constructed expert
reference labels independently for both organizational policy corpora.
The annotation followed the coverage definitions introduced in
Section~\ref{sec:assessment_framework} and the detailed rubric provided
in Appendix~\ref{appsec:anno_rubric}.

Two security researchers with seven and four years of cybersecurity
experience independently reviewed the controls. Both were familiar with
NIST SP~800-53 and had experience interpreting organizational security
policies in audit-related settings. For each organization, each
annotator reviewed all $1{,}007$ controls and assigned one of three
coverage labels:

\begin{itemize}
    \item \texttt{FULLY\_COVERED},
    \item \texttt{PARTIALLY\_COVERED}, or
    \item \texttt{NOT\_COVERED}.
\end{itemize}

The annotations were completed independently during the initial review.
Most disagreements occurred between \texttt{FULLY\_COVERED} and
\texttt{PARTIALLY\_COVERED}. These cases typically involved policies
that addressed the main security objective but left important details
unspecified. Common sources of disagreement included missing
responsibility assignments, unclear scope, absent review frequency,
limited procedural detail, and weak enforcement language.

The disagreements were subsequently reviewed through structured
adjudication. A control was treated as \texttt{FULLY\_COVERED} only when
the available policy documentation clearly addressed the control
requirement and its main governance elements. Cases in which relevant
policy evidence existed but important information remained incomplete
were assigned \texttt{PARTIALLY\_COVERED}. The resulting consensus
labels serve as the reference assessments used throughout the study.

Table~\ref{tab:label_distribution} shows the final distribution of
coverage labels. In both organizations, fully covered controls account
for a relatively small share of the total, while partial and absent
coverage make up the majority of the reference assessments. This
distribution also highlights why distinguishing partial coverage from
complete coverage is important in organizational policy review.

\begin{table}[htbp]
\centering
\begin{tabular}{l cc cc}
\toprule
\textbf{Coverage Label}
& \multicolumn{2}{c}{\textbf{OrgA}}
& \multicolumn{2}{c}{\textbf{OrgB}} \\
& Count & (\%) & Count & (\%) \\
\midrule
\texttt{FULLY\_COVERED}
& 110 & (10.9\%) & 130 & (12.9\%) \\

\texttt{PARTIALLY\_COVERED}
& 421 & (41.8\%) & 437 & (43.4\%) \\

\texttt{NOT\_COVERED}
& 476 & (47.3\%) & 440 & (43.7\%) \\

\midrule
\textbf{Total Controls}
& 1007 & (100\%) & 1007 & (100\%) \\
\bottomrule
\end{tabular}
\caption{Distribution of expert reference assessments across the two
organizational policy corpora after adjudication.}
\label{tab:label_distribution}
\end{table}

\paragraph{Inter-Annotator Agreement.}

We calculate Cohen's $\kappa$ on the independent annotations before
adjudication. Agreement is $\kappa=0.76$ for OrgA and $\kappa=0.71$ for
OrgB. Both values fall within the commonly used range for substantial
agreement.

The remaining disagreements illustrate an important characteristic of
cybersecurity policy assessment. The boundary between complete and
partial documentation can require professional judgment, particularly
when a policy addresses the intent of a control but leaves one or more
governance elements unspecified. The observed agreement nevertheless
shows that the two reviewers reached broadly consistent conclusions
before adjudication. Final consensus labels are used for all subsequent
analyses.

\subsection{Evaluation Measures}
\label{sec:metrics}

The evaluation considers five aspects of policy assessment: coverage
accuracy, security-relevant error patterns, evidence quality, the
usefulness of diagnostic outputs, and computational feasibility.

\paragraph{Coverage Assessment.}

Control-level assessments are compared with the expert reference labels
using accuracy, precision, recall, and F1-score. Because the study
contains three coverage categories with different frequencies, we
report class-sensitive performance in addition to overall accuracy.
Confusion matrices are used to examine where errors occur, particularly
at the boundary between full and partial coverage.






\paragraph{Evidence Quality.}

We separately evaluate whether the retrieved policy passages provide
useful evidence for the resulting assessment. This analysis uses a
stratified subset of 200 controls from each organization, with controls
selected across the three coverage categories.

Retrieval precision at rank $k$ is calculated as:

\begin{equation}
\label{eq:precision_at_k}
\text{Precision@k}
=
\frac{\text{Number of relevant excerpts in top-}k}{k}.
\end{equation}

An excerpt is treated as relevant when it contains information that can
reasonably contribute to assessment of the corresponding control under
the annotation rubric.

We also report the Top-1 sufficiency rate, which measures the proportion
of controls for which the highest-ranked excerpt alone provides enough
information to support the coverage assessment. This measure helps
distinguish between retrieval that merely finds related content and
retrieval that provides practically useful evidence.

\paragraph{Gap, Recommendation, and Explanation Quality.}

For controls assessed as \texttt{PARTIALLY\_COVERED} or
\texttt{NOT\_COVERED}, expert reviewers examine the additional
information produced by the assessment procedure.
Three dimensions are considered:

\begin{itemize}
    \item \textbf{Gap correctness:} whether the identified deficiencies
    accurately reflect elements missing from the policy relative to the
    control requirement;

    \item \textbf{Recommendation usefulness:} whether the suggested
    policy-level improvement is relevant, actionable, and consistent
    with the identified gap; and

    \item \textbf{Evidence groundedness:} whether the explanation is supported by the cited policy evidence and accurately connects the control requirement, coverage assessment, and identified gaps without introducing unsupported claims.
\end{itemize}

Each dimension is evaluated by expert reviewers using a five-point
Likert scale. The resulting scores provide an assessment of whether the
outputs are useful for policy review rather than simply whether the
coverage label is correct.

\paragraph{Computational Feasibility.}

We also record the computational resources required to perform the
assessment. These measures include the average number of tokens consumed
per control, average processing latency, and total monetary cost for a
complete organizational policy assessment. These results are used to
consider the practical feasibility of applying the approach to
large-scale policy review.

\subsection{Robustness Across Language Models}
\label{sec:models}

The primary objective of the study is to evaluate the policy-assessment
procedure rather than compare language models. Nevertheless, the quality
of an automated assessment may depend on the underlying model. We
therefore repeat the evaluation using several model families as a
robustness analysis.

We consider four hosted models---GPT-4o, GPT-4o-mini, Claude 3.5
Sonnet, and Claude 3.5 Haiku---and four locally deployed models---LLaMA
3.1 8B, Mistral 7B, Qwen2.5 7B, and
DeepSeek-R1-Distill-Llama-8B.

All models are evaluated under the same assessment procedure and use
the same prompt templates. No task-specific fine-tuning is performed.
The models receive no expert reference labels during inference, and the
original NIST control text is used consistently across runs. 

To reduce variation caused by sampling, the evaluations use a
temperature of 0 and a top-$p$ value of 1. Model outputs are compared
with the expert reference assessments only after inference is complete.

This analysis allows us to determine whether the observed findings are
specific to a particular model or remain visible across model families
with different capabilities and deployment settings.

\subsection{Comparison Conditions}
\label{sec:systems}

To understand which parts of the assessment procedure contribute to its
performance, we compare PACE with five alternative configurations.
The comparison varies three factors that are especially relevant to
policy assessment: whether evidence is retrieved before assessment, how
that evidence is retrieved, and whether coverage and gap analysis are
performed in separate stages.

For the main comparison, the same language-model backbone is used across
the configurations so that differences can be attributed to the
assessment procedure rather than to differences in model capability.

\paragraph{PACE.}

The complete assessment procedure uses semantic policy-evidence
retrieval followed by separate stages for coverage assessment, gap
identification, policy recommendation, and evidence-linked explanation.
Gap analysis and recommendation generation are performed only when the
available documentation is assessed as partial or absent.

\paragraph{B1: Direct Assessment Without Retrieval.}

The first comparison condition provides the security-control text and
policy corpus directly to the model and asks it to determine the
coverage category without a separate evidence-retrieval step. This
condition examines whether explicit evidence selection is necessary for
reliable control-level assessment.

\paragraph{B2: Semantic Retrieval with Single-Stage Assessment.}

The second condition retains semantic evidence retrieval but combines
coverage assessment, gap identification, and recommendation generation
within a single analysis step. This configuration is based on the
policy-analysis approach used in PARAG~\cite{saha2025parag} and allows
us to examine whether separating the assessment stages provides an
additional benefit beyond retrieval alone.

\paragraph{B3: Lexical Retrieval with Coverage Assessment.}

The third condition replaces semantic retrieval with BM25. The original
control text is used as the retrieval query, and the retrieved passages
are then used for coverage assessment. This condition tests whether
lexical similarity is sufficient for locating policy evidence when the
control and organizational policy use different terminology.

\paragraph{B4: Dense Retrieval with Single-Stage Assessment.}

The fourth condition uses standard embedding-based retrieval followed by
re-ranking and a single-stage coverage assessment. This provides a
stronger retrieval-based comparison while retaining a less structured
assessment procedure than PACE.

\paragraph{B5: Document-Level Assessment.}

The final condition retrieves complete documents rather than
fine-grained policy segments before performing the coverage assessment.
This comparison examines whether preserving large amounts of document
context provides an advantage or instead introduces unrelated material
that makes control-level assessment more difficult.

These comparison conditions are not intended to represent every
possible compliance-analysis technique. Rather, they isolate common
choices in automated policy assessment and allow us to examine the
contribution of evidence retrieval, retrieval granularity, and staged
analysis under otherwise comparable conditions.

\section{Results}
\label{sec:pro_results}

This section reports the results of the policy-assessment study across
the two organizational policy corpora. We first examine how consistently
documented control coverage can be assessed across different model
backbones. We then compare PACE with alternative assessment
configurations to examine the role of evidence retrieval and structured
coverage analysis. Finally, we consider how these design choices affect
the distinction between complete and incomplete policy coverage.

\subsection{Control-Level Policy Assessment}
\label{subsec:overall_performance}

We begin with the complete assessment procedure described in
Section~\ref{sec:pro_methodology}. Each of the $1{,}007$ NIST SP~800-53
controls is assessed independently against the policy corpus of OrgA and
OrgB. Performance is measured against the expert reference assessments
described in Section~\ref{sec:ground_truth}.

\subsubsection{Robustness Across Model Backbones}
\label{subsunsec:parag_pipeline}

Table~\ref{tab:overall_performance} reports the control-level assessment
results across the eight language-model backbones. The results show
substantial differences across models as well as some variation between
the two organizations.

\begin{table*}[t]
\centering
\begin{tabular}{l l cccc cccc}
\toprule
 & & \multicolumn{4}{c}{OrgA} & \multicolumn{4}{c}{OrgB} \\
\cmidrule(lr){3-6} \cmidrule(lr){7-10}
\textbf{Type} & \textbf{Model}
& Acc (\%) & Prec (\%) & Rec (\%) & F1 (\%)
& Acc (\%) & Prec (\%) & Rec (\%) & F1 (\%) \\
\midrule

\multirow{4}{*}{\textbf{Hosted}}
& GPT-4o
& 87.09 & 81.70 & 87.33 & 83.67
& 79.64 & 74.89 & 79.18 & 76.10 \\

& GPT-4o-mini
& 77.06 & 70.44 & 75.11 & 71.49
& 75.57 & 71.40 & 76.25 & 72.57 \\

& Sonnet
& \textbf{90.67} & \textbf{86.43} & \textbf{91.85} & \textbf{88.54}
& \textbf{84.91} & \textbf{80.81} & \textbf{84.84} & \textbf{82.31} \\

& Haiku
& 84.91 & 79.45 & 85.53 & 81.46
& 80.04 & 75.45 & 80.03 & 76.69 \\

\midrule

\multirow{4}{*}{\textbf{Local}}
& Llama
& 41.31 & 39.40 & 39.83 & 37.61
& 54.02 & 51.38 & 53.08 & 50.40 \\

& Mistral
& 51.24 & 48.54 & 50.51 & 47.06
& \textbf{70.61} & \textbf{66.04} & \textbf{69.75} & \textbf{66.58} \\

& Qwen
& \textbf{51.94} & \textbf{50.19} & \textbf{53.80} & \textbf{48.65}
& 68.42 & 64.24 & 68.07 & 64.67 \\

& DeepSeek
& 35.35 & 34.59 & 34.72 & 32.49
& 40.81 & 39.96 & 41.53 & 38.62 \\

\bottomrule
\end{tabular}
\caption{Control-level policy coverage assessment across model
backbones for OrgA and OrgB. Values are reported in \%.}
\label{tab:overall_performance}
\end{table*}

Among the hosted models, Sonnet produces the highest F1-score on both
organizations, reaching 88.54\% on OrgA and 82.31\% on OrgB. GPT-4o
also performs consistently across the two corpora, with F1-scores of
83.67\% and 76.10\%, while Haiku achieves 81.46\% and 76.69\%.
GPT-4o-mini performs somewhat lower on OrgA but shows a smaller reduction
on OrgB.

The locally deployed models show greater variation. On OrgA, Qwen
obtains the highest F1-score among these models at 48.65\%, closely
followed by Mistral at 47.06\%. On OrgB, both improve substantially,
with Mistral reaching 66.58\% and Qwen 64.67\%. Llama and DeepSeek
remain lower across both organizations.

Two observations are important here. First, assessment quality depends
strongly on the underlying model. The difference is especially visible
between the hosted and locally deployed models in the configurations
evaluated in this study. Second, performance also changes across the two
organizational corpora. Most hosted models perform lower on OrgB, while
several of the locally deployed models perform better on OrgB than on
OrgA.

These differences suggest that the difficulty of policy assessment
depends not only on the model, but also on how an organization structures
and expresses its security requirements. Policy terminology, document
organization, and the level of detail used to describe responsibilities
and procedures can all affect the alignment between control requirements
and organizational evidence. The results therefore support evaluating
an assessment approach on the policy environment in which it is intended
to be used rather than assuming that performance will transfer unchanged
across organizations.

\subsubsection{Comparison with Alternative Assessment Configurations}
\label{subsubsec:baseline_comparison}

We next compare the complete assessment procedure with the five
alternative configurations described in Section~\ref{sec:systems}.
The same model backbone is used for all configurations in this
comparison so that differences primarily reflect the assessment
procedure rather than model capability.

Table~\ref{tab:baseline_comparison} reports the results.

\begin{table*}[t]
\centering
\begin{tabularx}{\textwidth}{Xcccccccc}
\toprule
& \multicolumn{4}{c}{\textbf{OrgA}}
& \multicolumn{4}{c}{\textbf{OrgB}} \\
\cmidrule(lr){2-5} \cmidrule(lr){6-9}

\textbf{Configuration}
& \textbf{Acc (\%)} & \textbf{Prec (\%)} & \textbf{Rec (\%)} & \textbf{F1 (\%)}
& \textbf{Acc (\%)} & \textbf{Prec (\%)} & \textbf{Rec (\%)} & \textbf{F1 (\%)} \\

\midrule

\textbf{PACE}
& \textbf{90.67}
& \textbf{86.43}
& \textbf{91.85}
& \textbf{88.54}
& \textbf{84.91}
& \textbf{80.81}
& \textbf{84.84}
& \textbf{82.31} \\

\midrule

B1: Direct assessment
& 54.62 & 51.65 & 54.65 & 50.52
& 51.84 & 49.13 & 51.23 & 48.53 \\

B2: Semantic retrieval + single-stage assessment
& 70.21 & 64.12 & 68.32 & 64.91
& 68.72 & 64.83 & 69.20 & 65.37 \\

B3: BM25 retrieval + assessment
& 60.48 & 56.77 & 60.39 & 55.95
& 56.90 & 54.92 & 58.71 & 54.28 \\

B4: Dense retrieval + single-stage assessment
& 78.45 & 72.62 & 78.18 & 74.20
& 72.19 & 68.55 & 74.02 & 69.77 \\

B5: Document-level assessment
& 56.31 & 52.81 & 55.60 & 51.79
& 53.62 & 51.75 & 55.30 & 51.13 \\

\bottomrule
\end{tabularx}
\caption{Comparison of control-level policy assessment configurations
across the two organizational policy corpora.}
\label{tab:baseline_comparison}
\end{table*}

The complete PACE procedure achieves the highest F1-score among the
evaluated configurations on both organizations. On OrgA, it reaches
88.54\%, compared with 74.20\% for the strongest alternative
configuration, B4. This corresponds to a difference of 14.34 percentage
points. On OrgB, PACE reaches 82.31\%, compared with 69.77\% for
B4, a difference of 12.54 percentage points.

The direct-assessment configuration, B1, produces substantially lower
performance on both organizations, with F1-scores of 50.52\% and
48.53\%. In this setting, the model receives the control and policy
material without a separate step for selecting relevant evidence. The
result suggests that asking a model to reason directly over a large
policy collection makes it more difficult to identify the specific
information needed for a control-level judgment.

Adding semantic retrieval in B2 produces a clear improvement. F1 rises
to 64.91\% on OrgA and 65.37\% on OrgB. Lexical retrieval in B3 also
improves on direct assessment, but remains below the semantic-retrieval
configuration on both organizations. This difference is consistent with
the nature of cybersecurity policy documents, where standardized control
language and organizational policy language may describe similar
practices using different terms.

B4 provides the strongest alternative result. It combines dense
retrieval with re-ranking before assigning the coverage label and reaches
74.20\% F1 on OrgA and 69.77\% on OrgB. Its performance nevertheless
remains below the staged assessment used by PACE. This finding
suggests that retrieving relevant evidence is an important part of the
task, but that evidence still needs to be evaluated for completeness
before a coverage judgment is made.

The document-level configuration, B5, performs only slightly better than
the direct-assessment condition. Retrieving an entire policy document
provides more context, but it also introduces content that may have
little relevance to the specific control under review. Fine-grained
evidence retrieval appears better suited to identifying the particular
statements needed for control-level assessment.

\subsection{Role of Policy Evidence Retrieval}
\label{subsec:evidence_retrueve}

To examine the contribution of explicit policy evidence retrieval more
closely, we compare B1 and B2. These configurations differ primarily in
whether relevant policy segments are selected before the coverage
assessment.

Introducing semantic retrieval increases F1 from 50.52\% to 64.91\% on
OrgA and from 48.53\% to 65.37\% on OrgB. The corresponding improvements
are 14.39 and 16.84 percentage points, respectively.

This result indicates that evidence selection is important for
control-level policy assessment. Security policies often contain many
statements that are unrelated to the particular control being examined.
Identifying a smaller set of relevant passages gives the assessment a
more focused documentary basis.

The benefit is not limited to classification performance. Explicit
retrieval also preserves the connection between the resulting assessment
and the organization's policy documents. A reviewer can inspect which
passages were considered and determine whether they are relevant to the
control. This is particularly important because the assessment concerns
what the organization has formally documented. General cybersecurity
knowledge cannot substitute for policy evidence that is absent from the
organizational corpus.

At the same time, retrieval alone does not account for the performance
of the complete procedure. B2 remains 23.63 percentage points below
PACE on OrgA and 16.94 percentage points below it on OrgB. The next
analysis examines this difference.

\subsection{Role of Structured Coverage Assessment}
\label{subsec:staged_assessment}

PACE and B2 both use semantic retrieval, but they differ in how the
retrieved evidence is assessed. B2 asks for the coverage decision,
identified gaps, and recommendation within a single analysis step.
PACE first determines the coverage state and then performs gap and
recommendation analysis only when the evidence is incomplete or absent.

This change is associated with a substantial improvement in control-level
assessment. F1 increases from 64.91\% to 88.54\% on OrgA, a difference
of 23.63 percentage points. On OrgB, F1 increases from 65.37\% to
82.31\%, corresponding to a difference of 16.94 percentage points.

The distinction is particularly relevant for controls where the policy
addresses the general security topic but does not document all required
elements. For example, a policy may state that user access should be
reviewed periodically without specifying who is responsible for the
review, how frequently it should occur, or which accounts fall within
its scope. A single-stage assessment may identify the passage as
relevant and treat the control as fully addressed even though important
details remain unspecified.

Separating coverage assessment from gap identification makes this
boundary more explicit. The retrieved evidence is first examined for
whether it provides complete, partial, or no documented coverage.
When the result is partial, the subsequent analysis identifies the
specific elements that remain missing.

The results therefore point to an important distinction between
\emph{finding relevant policy evidence} and \emph{establishing that the
evidence is sufficient}. Semantic retrieval helps with the first task,
while the structured coverage assessment addresses the second. Both contribute to performance under the evaluated configuration

\subsection{Evidence Relevance and Sufficiency}
\label{subsec:evidence_quality}

The previous analyses show that explicit evidence retrieval improves
control-level assessment. We next examine whether the retrieved policy
passages are themselves relevant and sufficiently informative for
review.

This analysis uses the stratified, human-reviewed subset described in
Section~\ref{sec:metrics}. Table~\ref{tab:evidence_quality} reports
Precision@1, Precision@3, and the Top-1 sufficiency rate for both
organizations.

\begin{table}[t]
\centering
\begin{tabular}{lcc}
\toprule
\textbf{Metric} & \textbf{OrgA (\%)} & \textbf{OrgB (\%)} \\
\midrule
Precision@1 & 71.50 & 68.50 \\
Precision@3 & 83.50 & 80.00 \\
Top-1 Sufficiency Rate & 69.00 & 65.50 \\
\bottomrule
\end{tabular}
\caption{Relevance and sufficiency of retrieved policy evidence on the
human-reviewed subset.}
\label{tab:evidence_quality}
\end{table}

Precision@1 reaches 71.50\% for OrgA and 68.50\% for OrgB, showing that
the highest-ranked policy passage is relevant to the corresponding
control in most of the reviewed cases. When the first three retrieved
passages are considered, precision rises to 83.50\% for OrgA and
80.00\% for OrgB.

The Top-1 sufficiency rate is 69.00\% for OrgA and 65.50\% for OrgB.
Thus, in roughly two-thirds of the reviewed cases, the first retrieved
passage alone contains enough information to support the control-level
assessment. In the remaining cases, additional policy context is needed.

The difference between the Top-1 and Top-3 results is important for
organizational policy assessment. Relevant evidence is often distributed
across several sections or documents, and no single passage necessarily
captures the full requirement. The results therefore support retrieving
multiple candidate passages rather than assuming that the highest-ranked
excerpt is always sufficient.

They also reinforce the distinction between evidence relevance and
evidence completeness. A passage may be clearly related to a security
control while still omitting information needed to determine whether
the control is fully documented.

\subsection{Assessment Errors and Coverage Boundaries}
\label{subsec:error_analysis}

\begin{figure*}[t]
\centering
\begin{subfigure}{0.23\linewidth}
    \includegraphics[width=\linewidth]{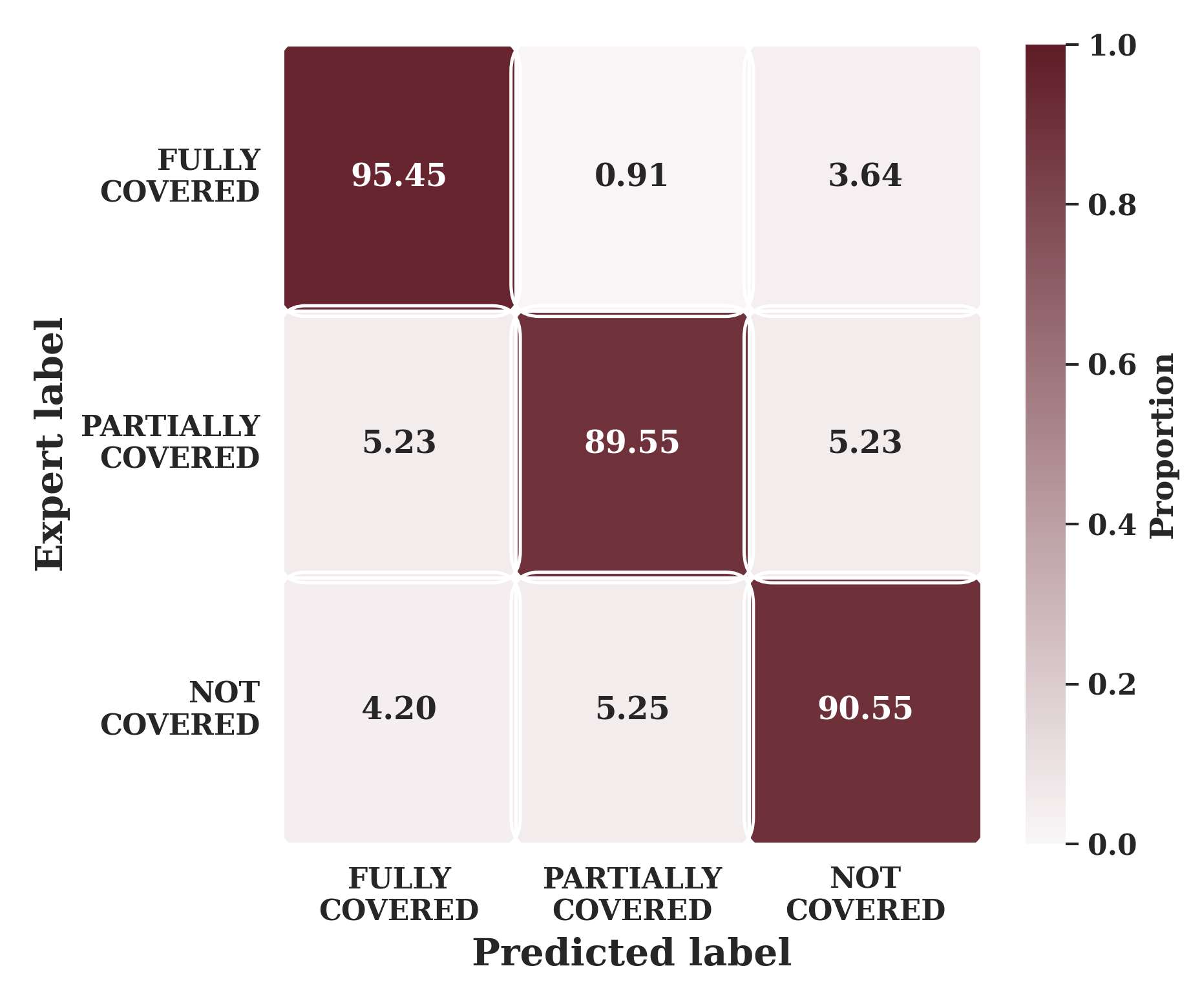}
    \caption{Sonnet}
\end{subfigure}
\begin{subfigure}{0.23\linewidth}
    \includegraphics[width=\linewidth]{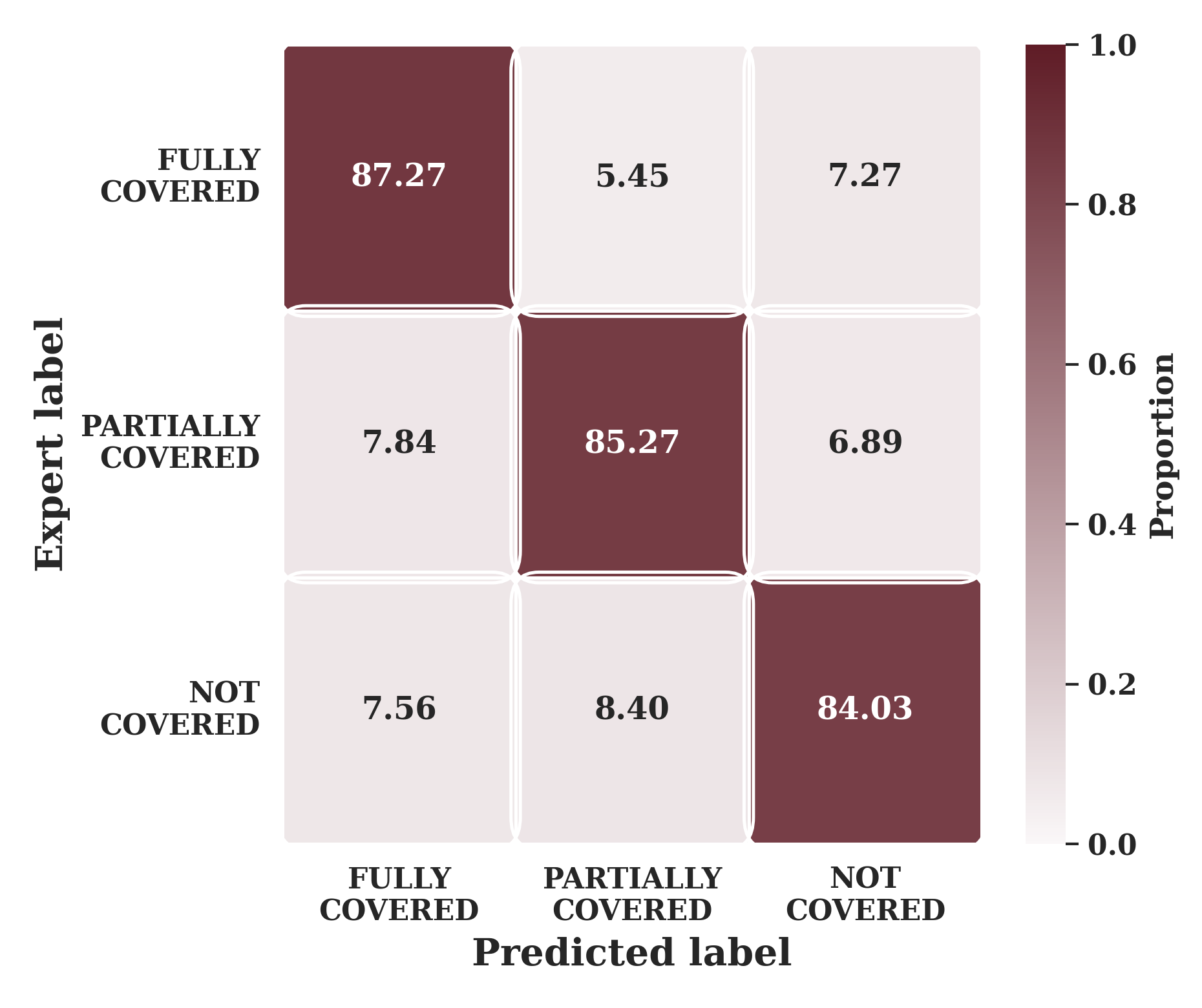}
    \caption{Haiku}
\end{subfigure}
\begin{subfigure}{0.23\linewidth}
    \includegraphics[width=\linewidth]{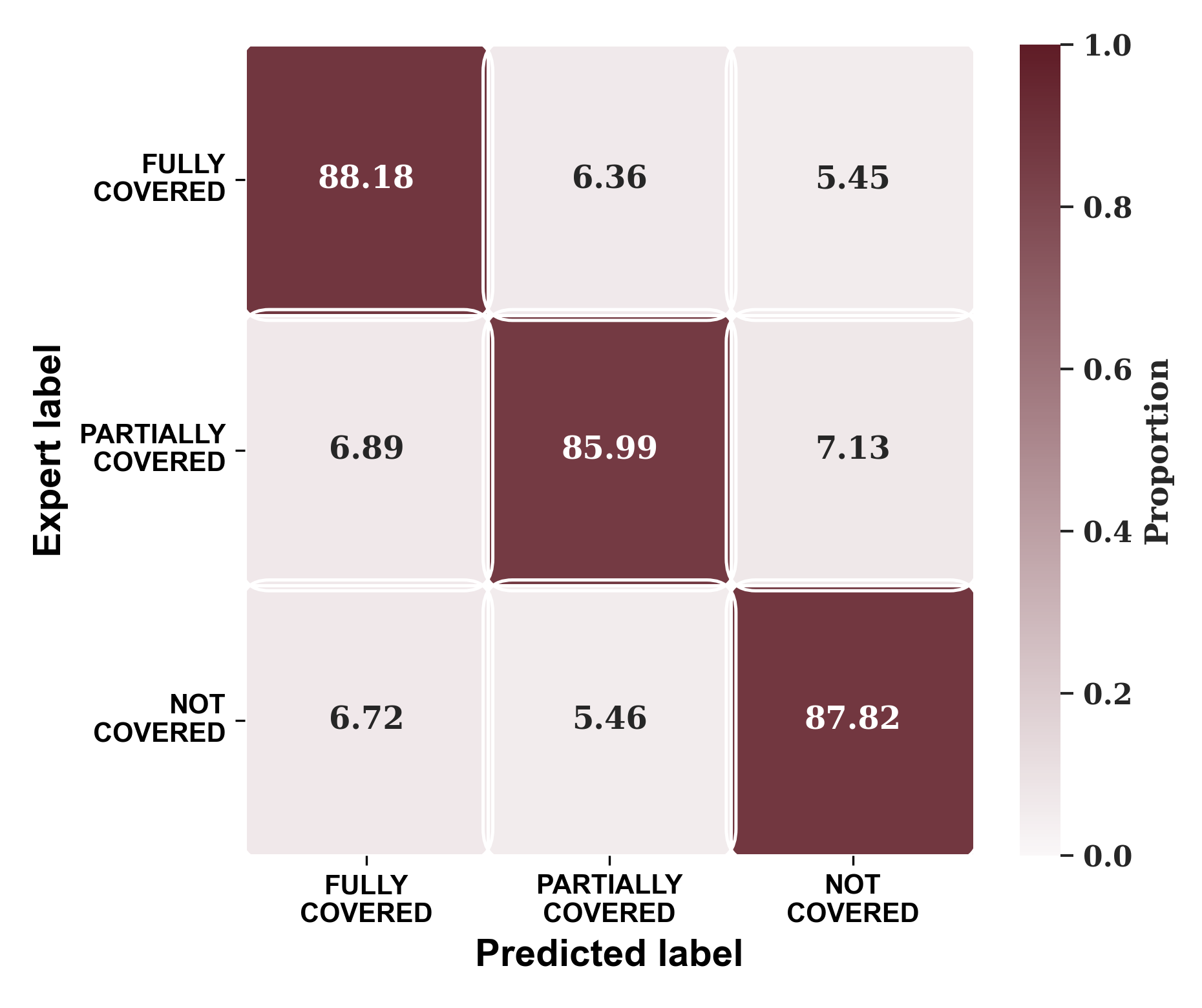}
    \caption{GPT-4o}
\end{subfigure}
\begin{subfigure}{0.23\linewidth}
    \includegraphics[width=\linewidth]{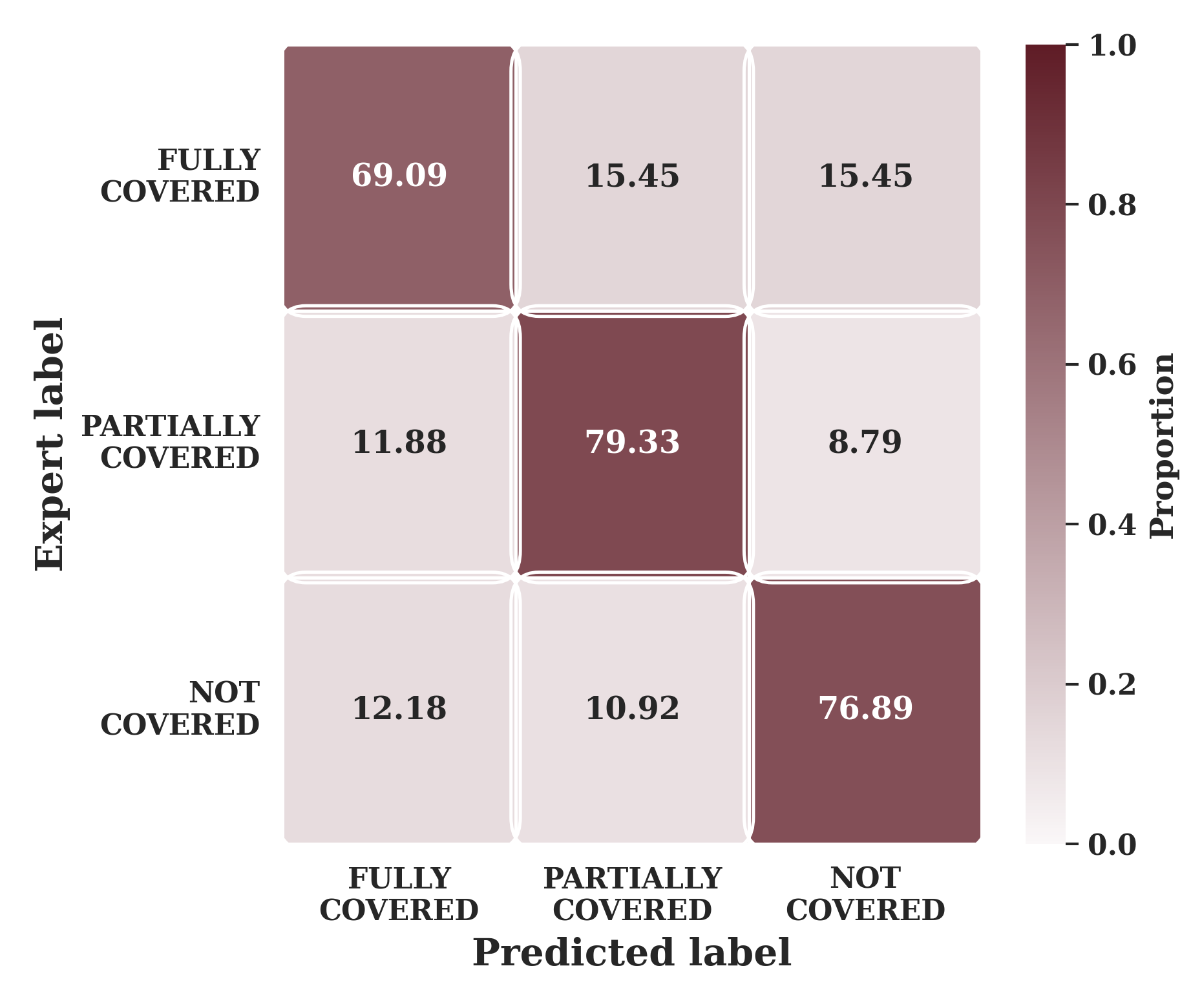}
    \caption{GPT-4o-mini}
\end{subfigure}
\begin{subfigure}{0.23\linewidth}
    \includegraphics[width=\linewidth]{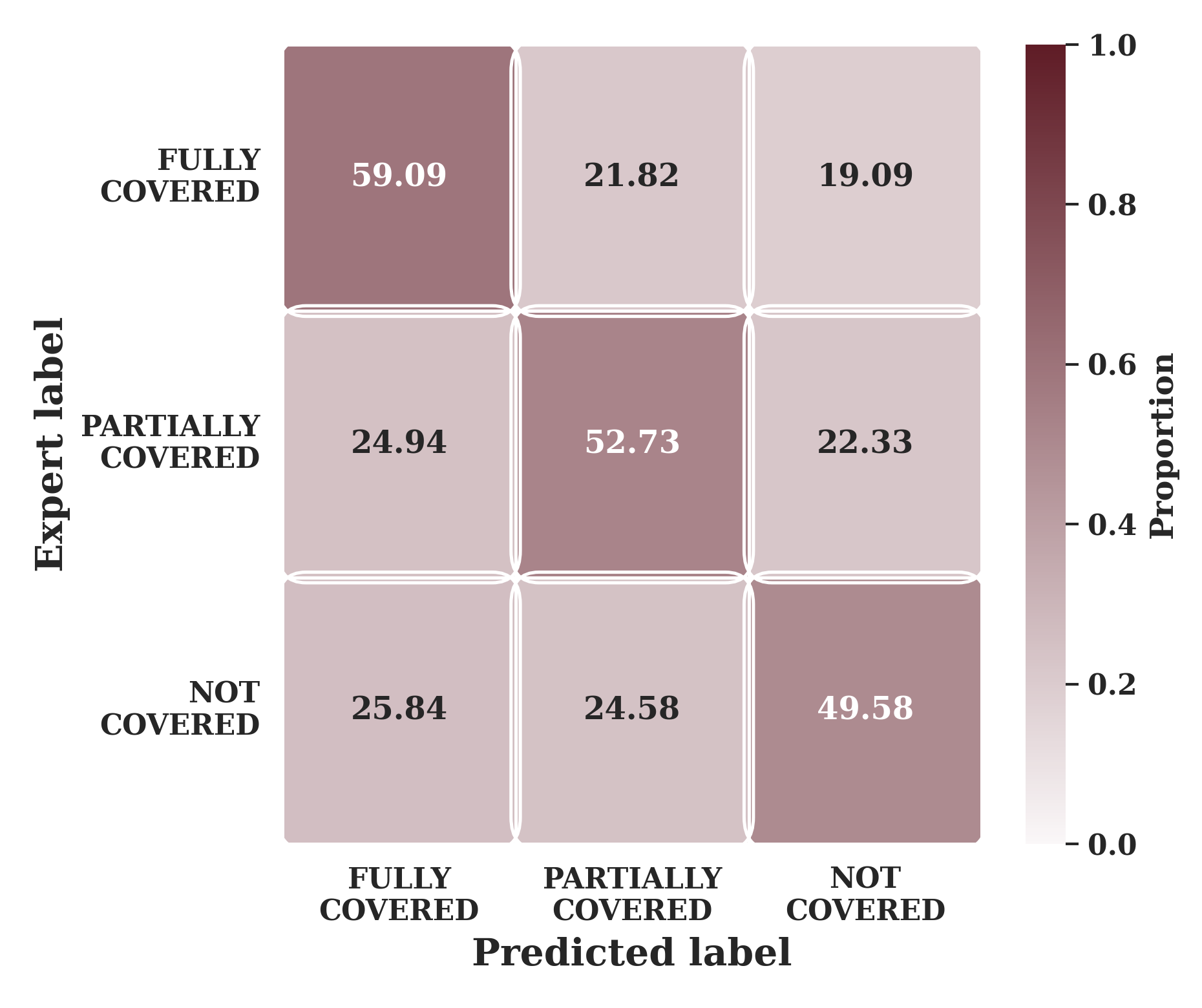}
    \caption{Qwen}
\end{subfigure}
\begin{subfigure}{0.23\linewidth}
    \includegraphics[width=\linewidth]{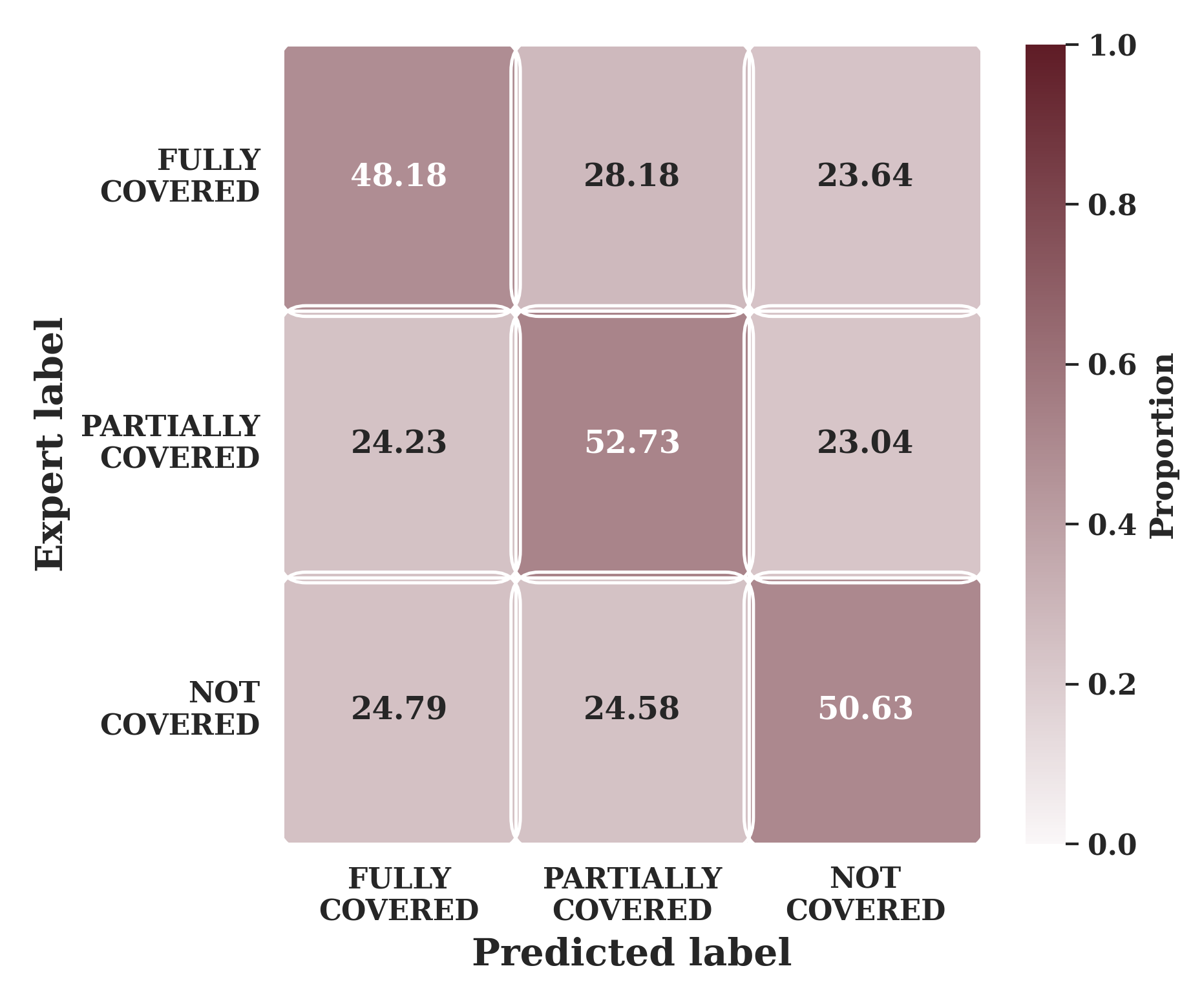}
    \caption{Mistral}
\end{subfigure}
\begin{subfigure}{0.23\linewidth}
    \includegraphics[width=\linewidth]{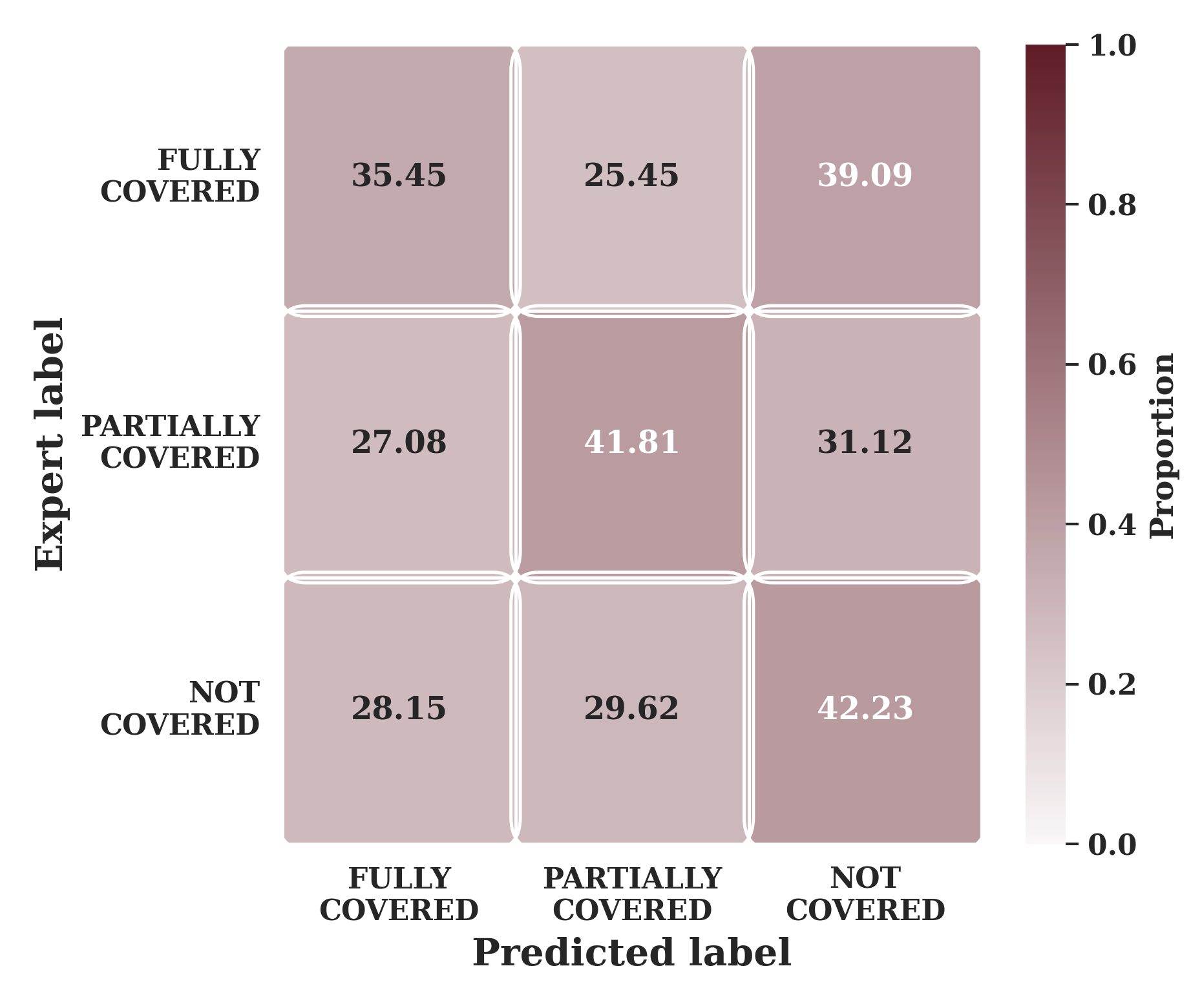}
    \caption{LLaMA}
\end{subfigure}
\begin{subfigure}{0.23\linewidth}
    \includegraphics[width=\linewidth]{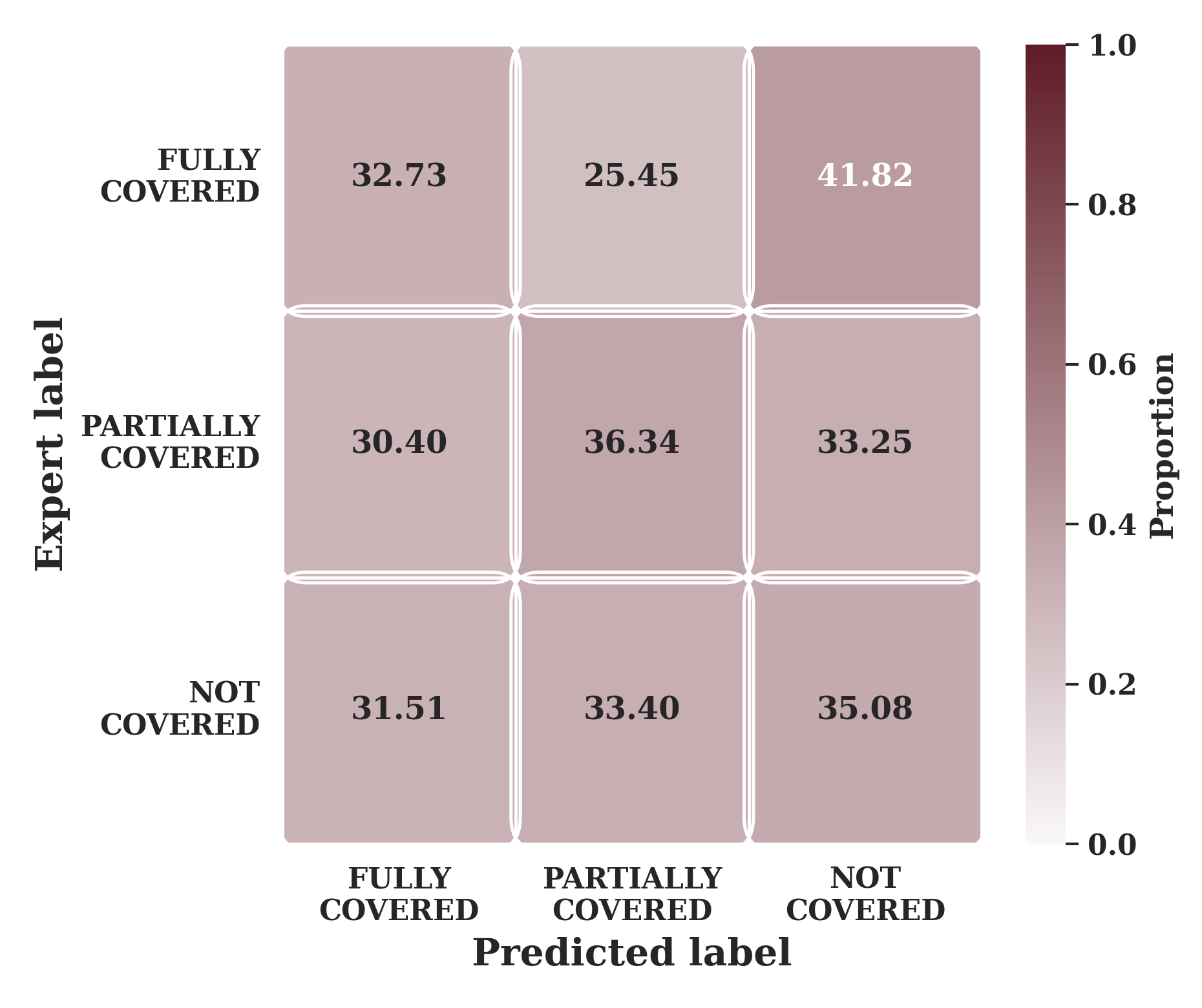}
    \caption{DeepSeek}
\end{subfigure}
\caption{Confusion matrices for control-level policy assessment on OrgA
across the evaluated model backbones.}
\label{fig:cm_orga}
\end{figure*}

\begin{figure*}[t]
\centering
\begin{subfigure}{0.23\linewidth}
    \includegraphics[width=\linewidth]{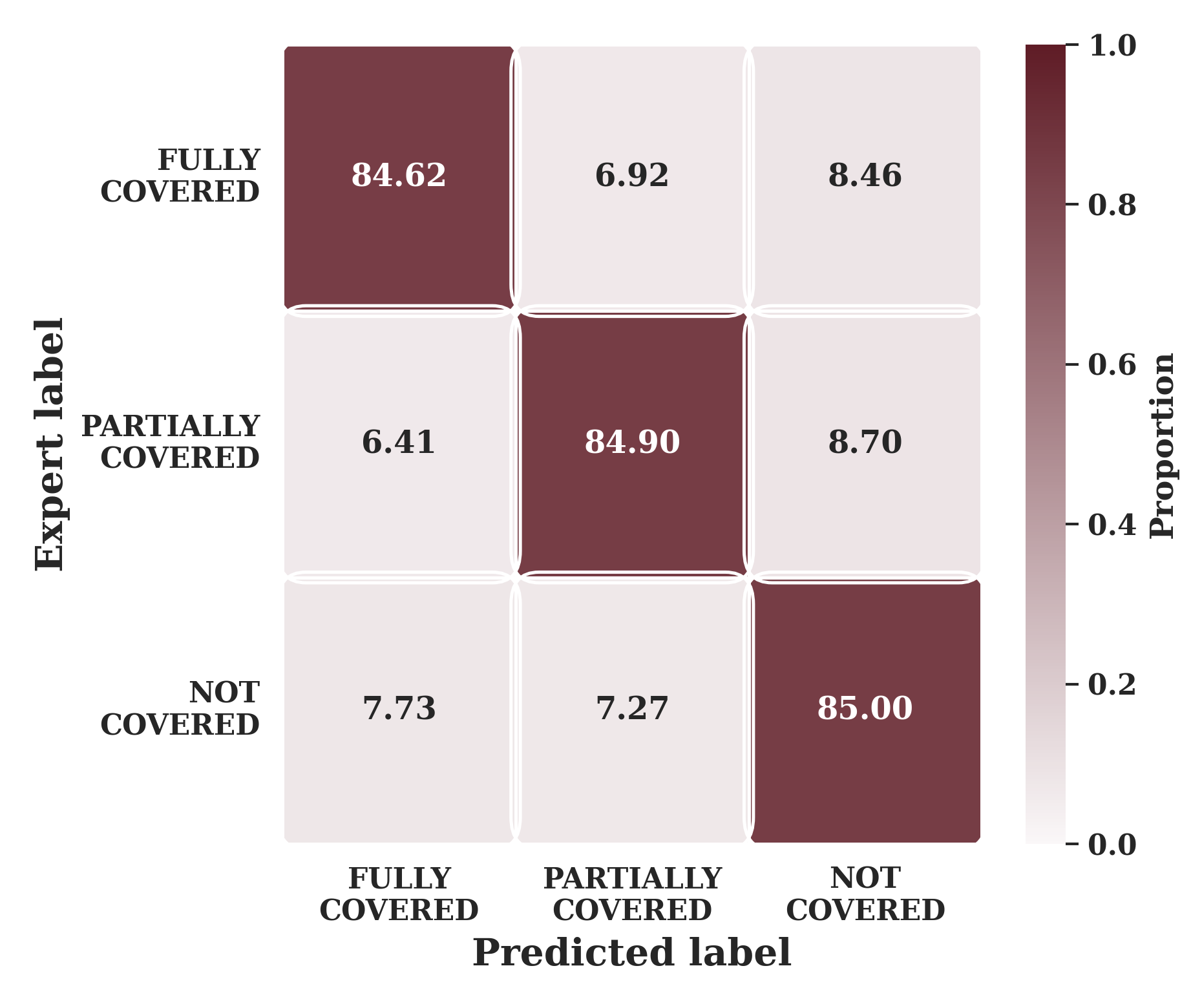}
    \caption{Sonnet}
\end{subfigure}
\begin{subfigure}{0.23\linewidth}
    \includegraphics[width=\linewidth]{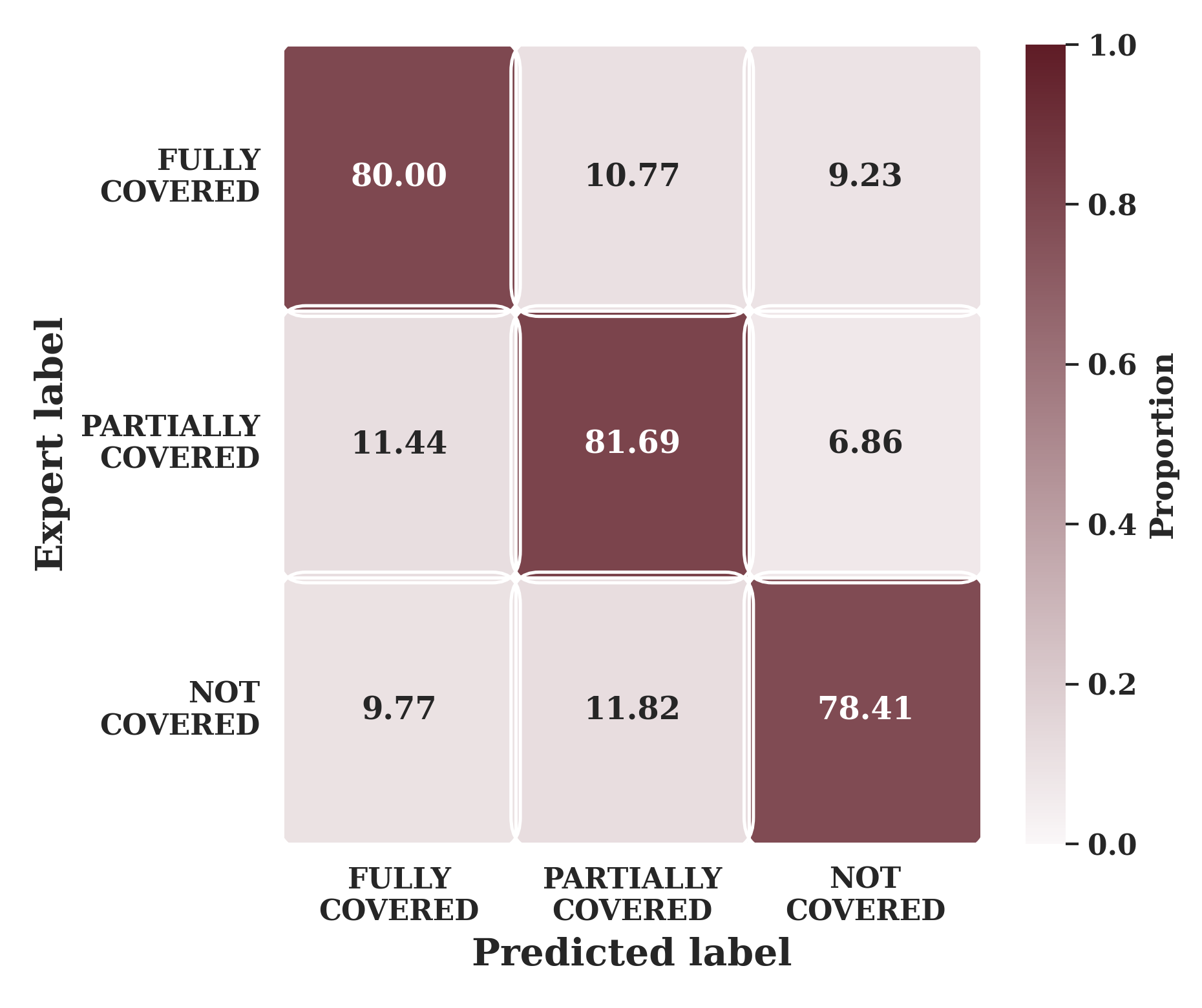}
    \caption{Haiku}
\end{subfigure}
\begin{subfigure}{0.23\linewidth}
    \includegraphics[width=\linewidth]{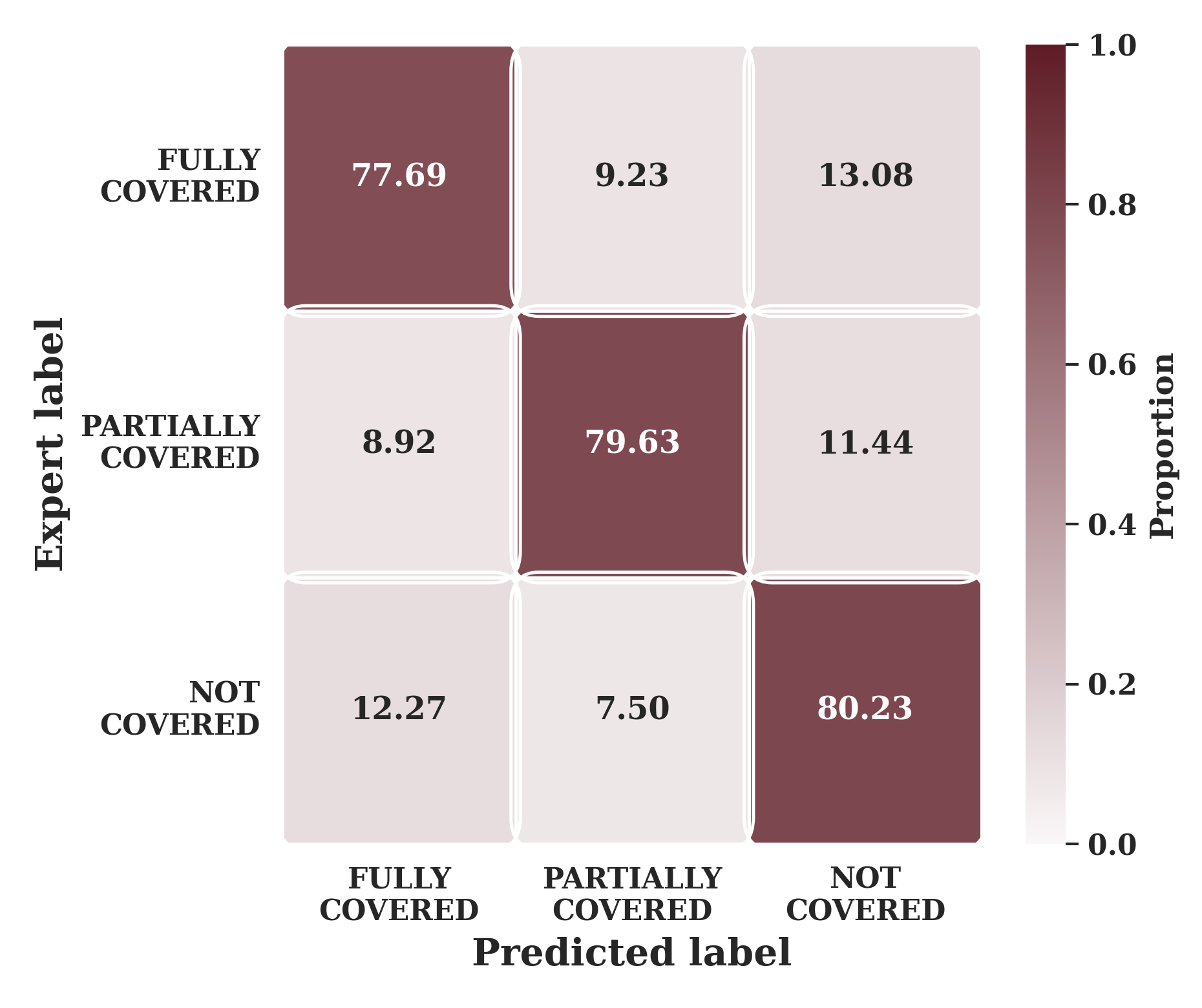}
    \caption{GPT-4o}
\end{subfigure}
\begin{subfigure}{0.23\linewidth}
    \includegraphics[width=\linewidth]{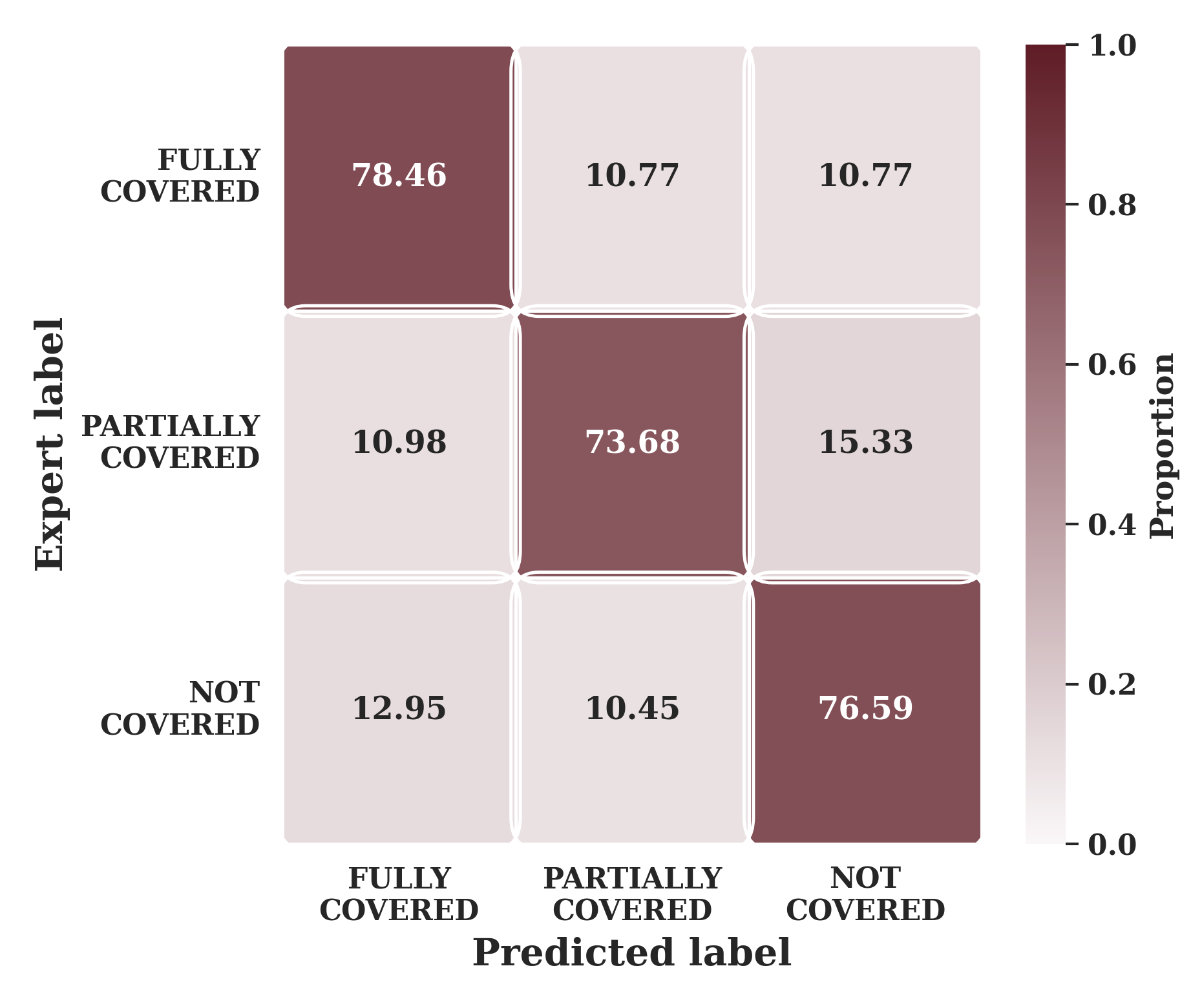}
    \caption{GPT-4o-mini}
\end{subfigure}
\begin{subfigure}{0.23\linewidth}
    \includegraphics[width=\linewidth]{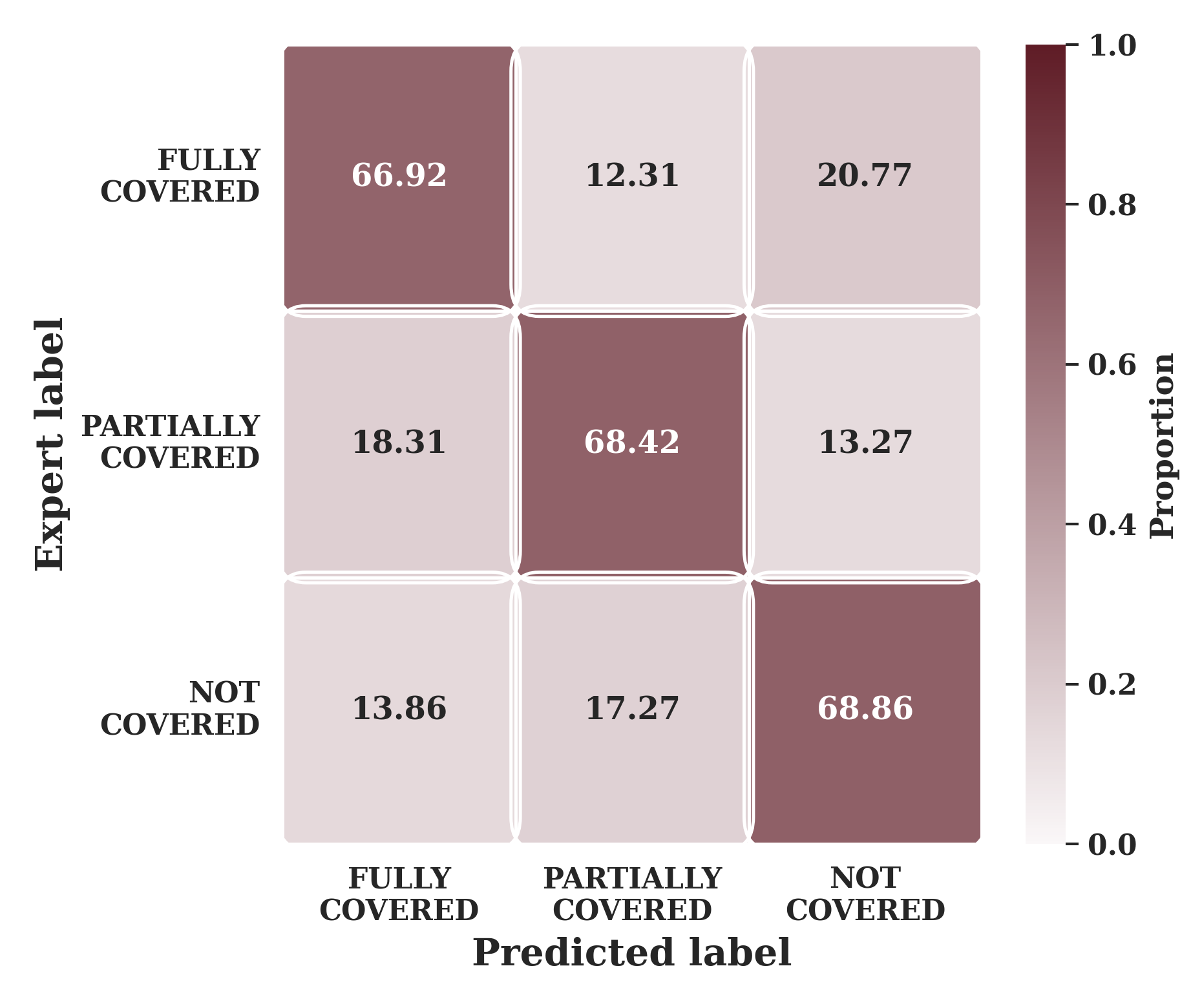}
    \caption{Qwen}
\end{subfigure}
\begin{subfigure}{0.23\linewidth}
    \includegraphics[width=\linewidth]{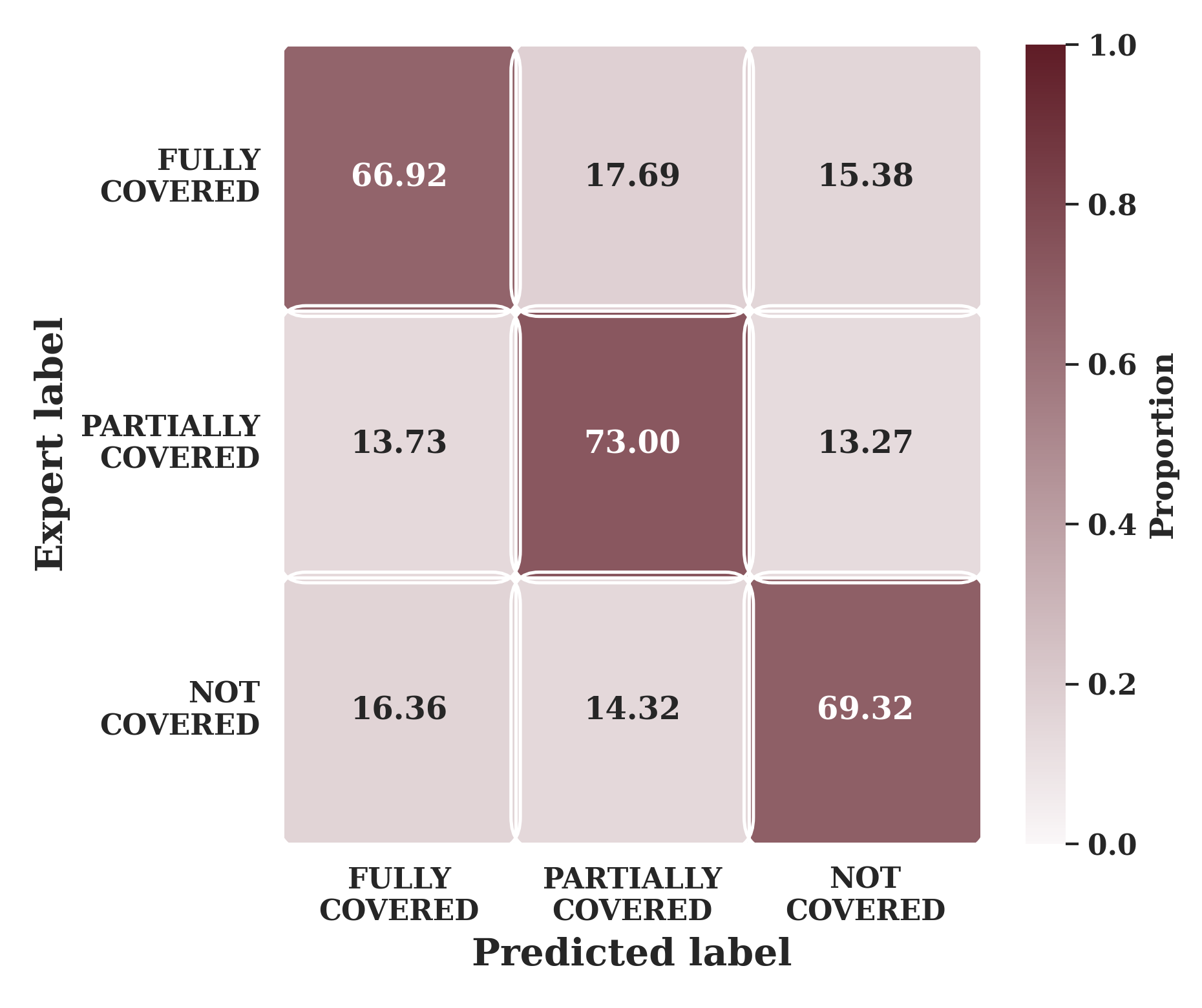}
    \caption{Mistral}
\end{subfigure}
\begin{subfigure}{0.23\linewidth}
    \includegraphics[width=\linewidth]{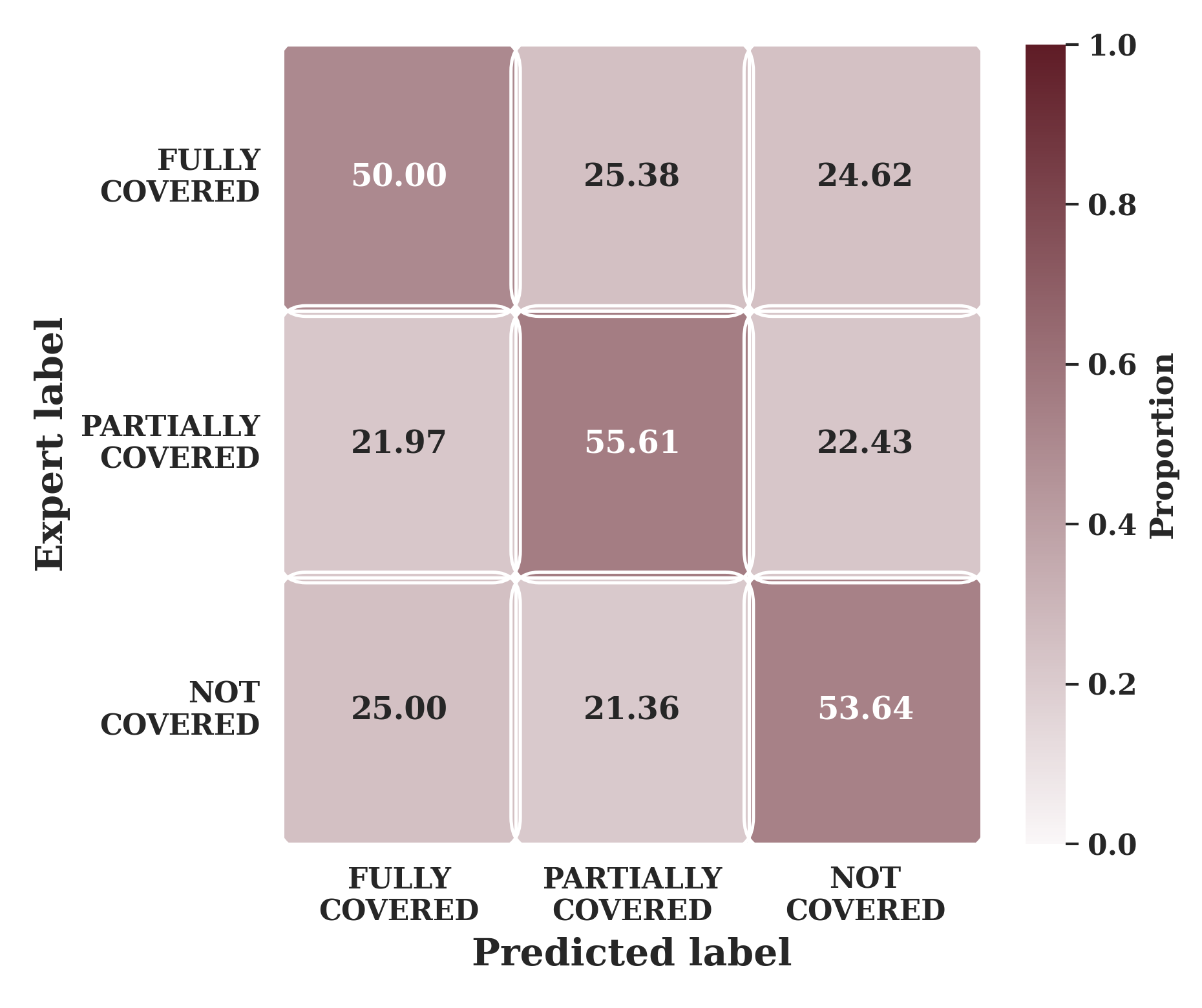}
    \caption{LLaMA}
\end{subfigure}
\begin{subfigure}{0.23\linewidth}
    \includegraphics[width=\linewidth]{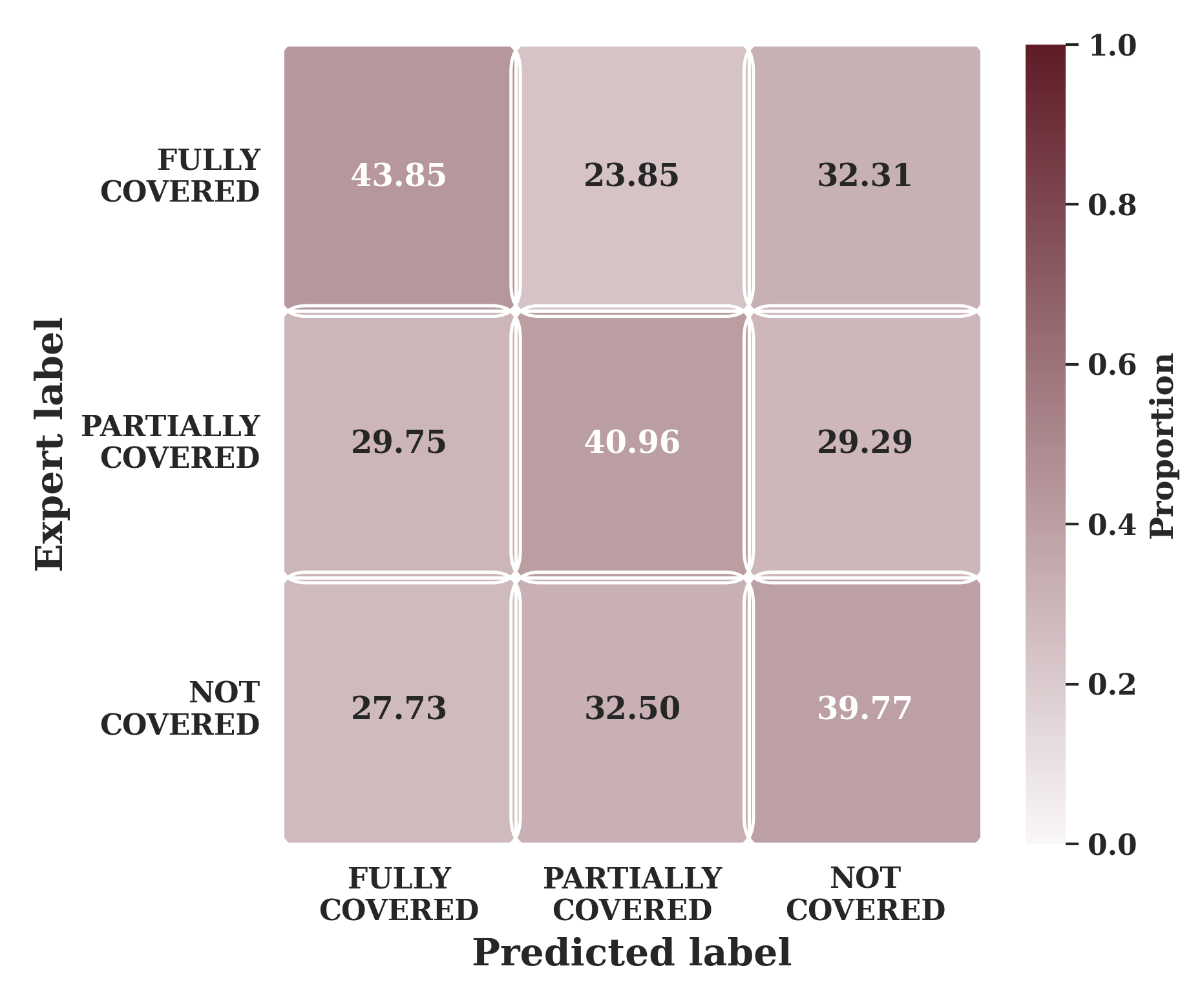}
    \caption{DeepSeek}
\end{subfigure}
\caption{Confusion matrices for control-level policy assessment on OrgB
across the evaluated model backbones.}
\label{fig:cm_orgb}
\end{figure*}

Aggregate accuracy and F1-scores do not show which coverage distinctions
are most difficult. We therefore examine the confusion matrices for the
eight model backbones across both organizations, shown in
Figures~\ref{fig:cm_orga} and~\ref{fig:cm_orgb}.

The confusion matrices show that the strongest-performing models produce
a clearer separation among the three coverage categories. Their remaining
errors are concentrated mainly at the boundary between
\texttt{FULLY\_COVERED} and \texttt{PARTIALLY\_COVERED}, or between
\texttt{PARTIALLY\_COVERED} and \texttt{NOT\_COVERED}. These are also the
cases in which the distinction depends most strongly on whether the
available documentation contains all of the required policy elements.

The \texttt{PARTIALLY\_COVERED} category is particularly important.
Partial coverage lies between two easier cases: documentation that
clearly addresses the control and documentation that provides little or
no relevant evidence. Determining partial coverage requires the reviewer
to recognize that relevant policy material exists while also identifying
what remains incomplete.

Errors involving this category can have different practical
consequences. Predicting \texttt{FULLY\_COVERED} when the expert
reference is \texttt{PARTIALLY\_COVERED} may create false assurance by
hiding a documentation gap. In the opposite direction, predicting
partial or absent coverage for a fully documented control can create
additional review or policy-revision work.

The lower-performing models show greater confusion across the three
categories, including cases where the predicted label differs
substantially from the expert reference assessment. This pattern is
consistent with the performance differences reported earlier and
further indicates that model choice influences the reliability of
control-level policy assessment.

At the same time, the presence of errors even among the stronger models
reinforces the need to preserve the supporting evidence and allow the
reviewer to inspect the suggested assessment rather than treating the
coverage label as an autonomous compliance decision.

\subsection{Gap and Recommendation Quality}
\label{subsec:diagnostic_quality}

For controls with partial or absent documented coverage, the usefulness
of the assessment depends on more than identifying the correct coverage
category. Reviewers also need to understand what is missing and what
policy changes may address the deficiency. We therefore evaluate the
quality of the identified gaps, policy recommendations, and
evidence-linked explanations.

Across the model outputs, recurring policy gaps include undefined
responsibility, missing procedures, absent review requirements, weak
enforcement language, and incomplete specification of the control
requirement. These categories correspond closely to the types of
deficiencies that distinguish partial documentation from complete
coverage.

Because the number and wording of generated gaps vary considerably
across model backbones, we focus the main analysis on expert judgments
of their quality rather than the raw number of gaps produced. 
Table~\ref{tab:diagnostic_quality} summarizes the expert evaluation.

\begin{table}[t]
\centering
\begin{tabular}{lcc}
\toprule
\textbf{Metric (1--5)} & \textbf{OrgA} & \textbf{OrgB} \\
\midrule
Mean Gap Correctness & 4.3 & 4.1 \\
Mean Recommendation Usefulness & 3.7 & 3.5 \\
Mean Evidence Groundedness & 4.4 & 4.2 \\
\bottomrule
\end{tabular}
\caption{Expert evaluation of identified policy gaps, recommendations,
and evidence grounding.}
\label{tab:diagnostic_quality}
\end{table}

Gap correctness receives mean ratings of 4.3 for OrgA and 4.1 for OrgB.
These scores indicate that the identified deficiencies generally match
the elements that expert reviewers considered missing from the
corresponding policies. Common examples include unclear responsibility,
missing procedures, absent review cycles, and insufficiently specified
governance requirements.

Recommendation usefulness receives lower, although still positive,
ratings of 3.7 for OrgA and 3.5 for OrgB. This difference is informative.
Identifying what is missing from a policy appears easier than proposing
a policy change that fully reflects the organization's own governance
context. Some recommendations therefore provide useful general
direction but may still require adjustment by a policy owner or auditor
before they can be adopted.

Evidence groundedness receives the highest ratings, at 4.4 for OrgA and
4.2 for OrgB. The expert judgments indicate that the assessment
explanations and identified deficiencies are usually connected to the
retrieved organizational evidence rather than being presented without
a documentary basis.

Taken together, these results suggest that the diagnostic outputs are
useful primarily as review support. They can help direct attention to
specific deficiencies, but the recommendations should not be interpreted
as ready-to-adopt policy decisions. Organization-specific knowledge and
professional review remain necessary before a suggested change is
incorporated into policy.

\subsection{Computational Feasibility}
\label{subsec:computational_efficiency}

Finally, we examine the computational requirements of applying the
assessment procedure to a complete organizational policy corpus.
Table~\ref{tab:efficiency} reports token consumption, processing time,
and estimated cost under the API-based configuration used in the study.

\begin{table}[t]
\centering
\begin{tabular}{lc}
\toprule
\textbf{Metric} & \textbf{Value} \\
\midrule
Average Tokens per Control & 5,200 \\
Average Time per Control & 14.8 seconds \\
Cost per 100 Controls & \$8.40 \\
Total Time (1,007 controls) & 4.14 hours \\
Total Cost (1,007 controls) & \$84.60 \\
\bottomrule
\end{tabular}
\caption{Computational requirements of the evaluated API-based
configuration.}
\label{tab:efficiency}
\end{table}

On average, the assessment of one control consumes approximately 5,200
tokens and requires 14.8 seconds under the evaluated configuration.
Assessing all $1{,}007$ controls requires approximately 4.14 hours of
sequential processing and an estimated API cost of \$84.60.

These figures provide an indication of the resources required for
large-scale policy assessment, although actual deployment cost and
latency will depend on the selected model, service provider, hardware,
and pricing structure. The controls are assessed independently, which
also makes parallel processing possible where the deployment environment
supports it.

The retrieval index can be retained when the same organizational policy
collection is assessed again. This may reduce preparation overhead for
periodic reviews or for reassessment after individual policy documents
are updated.

\subsection{Summary of Findings}
\label{subsec:results_summary}

The results provide four main observations:
(i) explicit policy-evidence retrieval substantially improves
control-level assessment compared with direct analysis of the policy
corpus. Semantic retrieval also performs better than lexical retrieval,
reflecting the terminology differences that commonly exist between
standardized security controls and organizational policies.
(ii) relevant evidence and sufficient evidence are not the same.
The improvement obtained through structured coverage assessment shows
that retrieved policy passages must still be examined for completeness
before they can support a full-coverage judgment.
(iii) the distinction between full and partial coverage remains one of
the most important sources of assessment error. This boundary matters
for security practice because overstating partial documentation may
create false assurance, while overly conservative assessments can
increase unnecessary review effort.
(iv) the expert evaluation shows that identified gaps are generally
consistent with professional judgments and that explanations are usually well grounded in organizational policy evidence. Recommendations are
also useful, but their lower ratings indicate that they often require
additional organizational context before being adopted.

These findings support the use of automated analysis as an
aid to cybersecurity policy review rather than as a substitute for
professional assessment.

\section{Discussion}
\label{sec:pro_discussion}

The results show that automated cybersecurity policy assessment can be
useful when the assessment remains closely tied to the organization's
own policy evidence. The main value is not simply the ability to assign
a coverage label. Policy reviewers also need to know which statements
support that label, whether the available evidence is complete, and what
remains missing when a control is only partially documented.

Across the two organizational policy collections, evidence retrieval and
structured coverage assessment both contribute to better agreement with
expert judgments. At the same time, the differences observed across
models and organizations show that automated assessment is not equally
reliable in every setting. These findings have implications for how such
methods should be used in cybersecurity policy review and governance.

\subsection{Relevant Evidence Is Not the Same as Sufficient Evidence}
\label{subsec:discussion_evidence}

One of the clearest findings of this study is the difference between
locating policy text that is relevant to a security control and
establishing that the control is fully documented. Semantic retrieval
substantially improves assessment performance over direct analysis and
lexical retrieval, showing the importance of first identifying the
policy material that is most closely related to the control under
review.

However, retrieval alone is not enough. The strongest retrieval-based
comparison remains below the complete assessment procedure, and the
human evaluation shows that additional context is often needed even
when the highest-ranked passage is relevant. This reflects the way
organizational policies are written in practice. A single statement may
describe the general security objective while responsibility, scope,
review frequency, enforcement, or procedural details appear elsewhere
or remain unspecified.

This distinction is important for cybersecurity policy assessment
because a relevant passage can easily be mistaken for evidence of
complete coverage. For example, a policy that requires access reviews
addresses the correct security topic, but the documentation may still
be incomplete if it does not specify who performs the review, how often
it occurs, or which accounts are included. Treating such cases as fully
covered may give reviewers more confidence in the policy than the
documentation supports.

The findings therefore suggest that policy assessment should separate
two questions: whether relevant evidence exists and whether that
evidence is sufficient for the control requirement. The first is
primarily an evidence-location problem, while the second requires an
assessment of completeness.

\subsection{Partial Coverage and the Risk of False Assurance}
\label{subsec:discussion_partial_coverage}

The distinction between \texttt{FULLY\_COVERED} and
\texttt{PARTIALLY\_COVERED} is particularly important. Both the expert
annotation process and the model error analysis show that this boundary
is more difficult than simply distinguishing relevant from irrelevant
policy material.

Partial coverage is common because organizational policies often state
a security expectation without documenting all of the elements needed
to support it. Missing ownership, unclear scope, absent review cycles,
weak enforcement language, and incomplete procedures appear repeatedly
among the identified deficiencies. These are not necessarily indications
that the organization has no security practice in place. Rather, they
show that the written policy does not provide complete evidence of the
requirement being assessed.

From an audit and governance perspective, errors at this boundary do not
all have the same consequence. Predicting full coverage where the expert
assessment identifies only partial or absent coverage can create
\emph{false assurance}: a reviewer may conclude that no further policy
attention is needed even though important elements remain undocumented.
The opposite error is less likely to hide a policy deficiency but may
create unnecessary review or revision work.

This asymmetry is important when automated assessment is used in
practice. Aggregate accuracy or F1-score alone cannot show whether the
remaining errors tend to overstate or understate policy completeness.
For this reason, security-oriented evaluation should pay particular
attention to false-assurance errors and to cases near the boundary
between full and partial coverage.

\subsection{Value of Gap Identification for Policy Review}
\label{subsec:discussion_gaps}

A coverage label is more useful when the reviewer can understand why
the policy is incomplete. The expert evaluation indicates that the
identified gaps generally correspond well with the deficiencies
recognized by human reviewers. This is important because the practical
question following a partial or absent assessment is usually not simply
``Is this control covered?'' but ``What is missing from the policy?''

The recurring gap categories provide a useful description of the kinds
of documentation weaknesses that appear across the two policy
collections. These include unclear responsibility, missing procedures,
undefined review requirements, weak enforcement provisions, and
insufficiently specified control expectations. Presenting these
deficiencies explicitly can help a reviewer focus on the part of the
policy that requires clarification rather than re-examining the entire
document collection.

The recommendation results should be interpreted more cautiously.
Although expert reviewers generally considered the recommendations
useful, their ratings were lower than those for gap correctness and
evidence groundedness. This difference is reasonable because identifying
a missing policy element does not necessarily determine the best way to
address it. An appropriate policy change may depend on organizational
structure, existing responsibilities, legal requirements, risk appetite,
or technical practices that are not visible in the policy corpus.

Recommendations should therefore be treated as starting points for
policy revision rather than as text that can be adopted without further
review. The policy owner or security professional still needs to decide
how an identified deficiency should be addressed within the
organization's own governance context.

\subsection{Evidence Traceability and Professional Review}
\label{subsec:discussion_traceability}

Evidence traceability is important because the assessment concerns what
the organization has formally documented. A model may possess general
knowledge about common cybersecurity practices, but that knowledge
cannot serve as evidence that a particular organization has established
those practices in policy.

Retaining the document, section, and source location associated with
each retrieved passage allows the reviewer to return to the original
material and check whether the suggested interpretation is reasonable.
The high groundedness scores observed in the expert evaluation suggest
that this connection is generally preserved in the assessments produced
in our study.

This form of traceability also changes how automated outputs can be
used. Rather than asking a reviewer to accept a coverage label on its
own, the assessment presents a claim together with the documentary
material that supports it. The reviewer can then accept, revise, or
reject the suggested assessment.

Such review remains necessary even for the stronger-performing models.
The differences observed across model backbones and between OrgA and
OrgB show that assessment quality depends partly on the underlying model
and on characteristics of the policy corpus. An approach that performs
well for one organization should therefore not be assumed to provide the
same level of reliability in another organization without local
evaluation.

\subsection{Implications for Cybersecurity Policy Management}
\label{subsec:organizational_implications}

The findings suggest several practical uses for evidence-grounded policy
assessment.

First, automated retrieval can reduce the effort required to locate
relevant material in large policy collections. A reviewer assessing
hundreds of controls may otherwise need to search repeatedly across
multiple documents and reconcile different terminology. Presenting a
small set of candidate passages can make this part of the review more
manageable.

Second, the use of explicit coverage categories can support more
consistent treatment of incomplete documentation. In particular, a
separate partial-coverage category avoids forcing every control into a
binary covered/not-covered judgment. This is useful in policy
collections where requirements are frequently addressed in part but not
in full.

Third, recurring gap categories can help policy owners identify broader
documentation weaknesses. If missing ownership assignments, review
cycles, or enforcement provisions occur repeatedly across control
families, the issue may reflect a wider policy-writing or governance
problem rather than an isolated control deficiency.

Fourth, the relatively modest computational requirements observed in
our evaluation suggest that automated assessment could be used not only
for one-time reviews but also for reassessment after substantial policy
changes. For example, organizations could re-examine documented coverage
after revisions to security policies or updates to the control framework.
Such repeated use would require clear versioning of both the policy
corpus and the reference framework so that findings remain reproducible.

These potential uses should not be interpreted as evidence that the
approach reduces audit time or improves organizational security
outcomes. The present study evaluates the quality of policy assessments,
retrieved evidence, gaps, and recommendations. It does not measure how
much reviewer effort is saved in practice, whether organizations adopt
the suggested changes, or whether those changes improve security
performance. These questions require separate organizational studies.

\subsection{Conditions for Responsible Use}
\label{subsec:responsible_use}

The results also suggest several conditions that should be considered
before automated policy assessment is incorporated into operational
security or audit processes.

First, the outputs should remain advisory. The assessment is based on
written policy evidence and may miss organizational context that is
known to the reviewer but not captured in the documents.

Second, organizations should validate the selected model and assessment
procedure on their own policy material. The variation observed across
models and between the two organizational corpora indicates that
performance cannot be assumed to transfer unchanged to a new policy
environment.

Third, the evidence underlying an assessment should remain available to
the reviewer. A coverage label without its supporting source material
provides little basis for verification and increases the risk that an
unsupported conclusion is accepted simply because it was produced
automatically.

Finally, policy coverage should not be interpreted as evidence of
operational control effectiveness. PACE assesses whether a security
practice is sufficiently documented in organizational policy. It does
not determine whether that practice has been implemented correctly,
whether employees follow it, or whether the control performs effectively
in practice. Operational compliance would require additional evidence,
such as technical configurations, logs, audit records, interviews, or
observations of organizational practice.

\subsection{Limitations and Future Research}
\label{subsec:discussion_limitations}

Several limitations should be considered when interpreting the results.

First, the study uses policy collections from two organizations.
Although the two corpora differ in document structure and writing style,
they cannot represent the full range of cybersecurity policies used
across industries, organization sizes, jurisdictions, and regulatory
settings. Evaluation across a larger and more diverse set of
organizations would provide a stronger basis for assessing
generalizability.

Second, the study focuses on NIST SP~800-53. The assessment procedure
may also be applicable to other standards and regulatory frameworks, but
requirements differ in structure, terminology, and level of detail.
Future work should examine whether the same approach transfers to
frameworks such as ISO/IEC~27001 or sector-specific regulatory
requirements.

Third, policy assessment remains partly judgment based. The
inter-annotator agreement was substantial, but disagreements still
occurred, particularly between full and partial coverage. This reflects
a genuine ambiguity in cases where a policy captures the general intent
of a control but leaves some requirements implicit or incomplete.
Automated results should therefore be interpreted against the same
uncertainty that exists in human assessment.

Fourth, the expert evaluation of evidence, gaps, and recommendations was
conducted on a subset of the complete assessment output. Although this
provides direct human validation, broader evaluation involving
practicing auditors, compliance professionals, and organizational policy
owners would provide additional evidence about practical usefulness.

Fifth, model performance may change as model providers update their
systems, and locally deployed models may behave differently under other
prompting or inference configurations. The results should therefore be
understood as measurements of the evaluated configurations rather than
fixed properties of the model families.

Finally, the study evaluates documented policy coverage rather than
actual security implementation. A fully documented control may still be
poorly implemented, while a security practice may exist even when the
corresponding policy documentation is incomplete. Future studies could
combine policy evidence with implementation evidence to examine the
relationship between documented requirements and operational practice.

Future work can also examine cross-policy dependencies, assessments
against multiple control frameworks, and methods for expressing
uncertainty when the available evidence does not support a clear
coverage judgment. A particularly useful direction would be to study
the approach in an operational review setting and measure how it affects
reviewer effort, disagreement, decision time, and the quality of final
policy revisions.

\section{Conclusion}
\label{sec:conclusion}

This study examined how evidence-grounded automated analysis can support cybersecurity policy assessment. Using 1,007 NIST SP~800-53 controls and two real-world organizational policy corpora, we found that semantic evidence retrieval and structured coverage assessment improve the ability to distinguish full, partial, and absent policy coverage. Expert evaluation further shows that the identified gaps are generally consistent with human judgments and remain well grounded in organizational policy evidence. At the same time, partial coverage remains difficult to assess, and performance varies across models and organizations. These findings support the use of automated analysis as an aid to cybersecurity policy review, with source evidence kept visible and final judgments left to security professionals. Future work should examine the approach across additional organizations, security frameworks, and operational audit settings.

\section*{Declaration of Generative AI and AI-assisted technologies in the writing process}
During the preparation of this manuscript, the authors used AI-based tools, including ChatGPT, Gemini, and Claude, to assist with text refinement, sentence rewriting, and grammar and spelling corrections. All outputs generated by these tools were carefully reviewed and edited by the authors. The authors take full responsibility for the content and integrity of the published work.


\printcredits

\bibliographystyle{cas-model2-names}

\bibliography{cas-refs}

@article{knapp2009information,
  title={Information security policy: An organizational-level process model},
  author={Knapp, Kenneth J and Morris Jr, R Franklin and Marshall, Thomas E and Byrd, Terry Anthony},
  journal={Computers \& security},
  volume={28},
  number={7},
  pages={493--508},
  year={2009},
  publisher={Elsevier}
}

@article{cram2017organizational,
  title={Organizational information security policies: a review and research framework},
  author={Cram, W Alec and Proudfoot, Jeffrey G and D’arcy, John},
  journal={European Journal of Information Systems},
  volume={26},
  number={6},
  pages={605--641},
  year={2017},
  publisher={Taylor \& Francis}
}

@inproceedings{reimers2019sentencebert,
  title = "Sentence-BERT: Sentence Embeddings using Siamese BERT-Networks",
  author = "Reimers, Nils and Gurevych, Iryna",
  booktitle = "Proceedings of the 2019 Conference on Empirical Methods in Natural Language Processing",
  month = "11",
  year = "2019",
  publisher = "Association for Computational Linguistics",
  url = "https://arxiv.org/abs/1908.10084",
}

@inproceedings{mandal2018automating,
  title={Automating information security policy compliance checking},
  author={Mandal, Debashis and Mazumdar, Chandan},
  booktitle={2018 Fifth International Conference on Emerging Applications of Information Technology (EAIT)},
  pages={1--4},
  year={2018},
  organization={IEEE}
}

@techreport{force2020security,
  title={Security and privacy controls for information systems and organizations},
  author={Force, Joint Task},
  year={2020},
  institution={National Institute of Standards and Technology}
}

@article{force2022assessing,
  title={Assessing security and privacy controls in information systems and organizations},
  author={Force, Joint Task},
  journal={NIST Special Publication},
  volume={800},
  pages={53A},
  year={2022}
}

@inproceedings{xiao2012automated,
  title={Automated extraction of security policies from natural-language software documents},
  author={Xiao, Xusheng and Paradkar, Amit and Thummalapenta, Suresh and Xie, Tao},
  booktitle={Proceedings of the ACM SIGSOFT 20th International Symposium on the Foundations of Software Engineering},
  pages={1--11},
  year={2012}
}

@article{breaux2008analyzing,
  title={Analyzing regulatory rules for privacy and security requirements},
  author={Breaux, Travis and Ant{\'o}n, Annie},
  journal={IEEE transactions on software engineering},
  volume={34},
  number={1},
  pages={5--20},
  year={2008},
  publisher={IEEE}
}

@inproceedings{saha2025parag,
  title={PARAG: Proactive A nswering Framework Integrating LLMs with R etrieval-A ugmented G eneration},
  author={Saha, Bikash and Rani, Nanda and Chakraborty, Joheen and Singh, Divyanshu and Chakraborty, Soumyo V and Shukla, Sandeep Kumar},
  booktitle={European Interdisciplinary Cybersecurity Conference},
  pages={20--37},
  year={2025},
  organization={Springer}
}

@article{rodriguez2024large,
  title={Large language models: a new approach for privacy policy analysis at scale},
  author={Rodriguez, David and Yang, Ian and Del Alamo, Jose M and Sadeh, Norman},
  journal={Computing},
  volume={106},
  number={12},
  pages={3879--3903},
  year={2024},
  publisher={Springer}
}

@inproceedings{jain2025complexity,
  title={From Complexity to Clarity: AI/NLP’s Role in Regulatory Compliance},
  author={Jain, Jivitesh and Dhanasekaran, Nivedhitha and Diab, Mona},
  booktitle={Findings of the Association for Computational Linguistics: ACL 2025},
  pages={26629--26641},
  year={2025}
}

@article{slapnivcar2022effectiveness,
  title={Effectiveness of cybersecurity audit},
  author={Slapni{\v{c}}ar, Sergeja and Vuko, Tina and {\v{C}}ular, Marko and Dra{\v{s}}{\v{c}}ek, Matej},
  journal={International Journal of Accounting Information Systems},
  volume={44},
  pages={100548},
  year={2022},
  publisher={Elsevier}
}

@article{madan2025argen,
  title={ArGen: Auto-Regulation of Generative AI via GRPO and Policy-as-Code},
  author={Madan, Kapil},
  journal={arXiv preprint arXiv:2509.07006},
  year={2025}
}

@article{kholkar2025policy,
  title={Policy-as-prompt: Turning AI governance rules into guardrails for AI agents},
  author={Kholkar, Gauri and Ahuja, Ratinder},
  journal={arXiv preprint arXiv:2509.23994},
  year={2025}
}

@inproceedings{romeo2025arpaccino,
  title={Arpaccino: an agentic-rag for policy as code compliance},
  author={Romeo, Francesco and Arena, Luigi and Blefari, Francesco and Pironti, Francesco Aurelio and Lupinacci, Matteo and Furfaro, Angelo},
  booktitle={European Conference on Advances in Databases and Information Systems},
  pages={467--481},
  year={2025},
  organization={Springer}
}

@article{tummala2026autonomous,
  title={Autonomous agents and policy compliance: A framework for reasoning about penalties},
  author={Tummala, Vineel and Inclezan, Daniela},
  journal={Theory and Practice of Logic Programming},
  pages={1--30},
  year={2026},
  publisher={Cambridge University Press}
}

@article{agarwal2025five,
  title={A five-layer framework for AI governance: integrating regulation, standards, and certification},
  author={Agarwal, Avinash and Nene, Manisha J},
  journal={Transforming Government: People, Process and Policy},
  volume={19},
  number={3},
  pages={535--555},
  year={2025},
  publisher={Emerald Publishing Limited}
}

@book{edwards2024comprehensive,
  title={A comprehensive guide to the NIST cybersecurity framework 2.0: Strategies, implementation, and best practice},
  author={Edwards, Jason},
  year={2024},
  publisher={John Wiley \& Sons}
}

@INPROCEEDINGS{hassani2024rethinking,
  author={Hassani, Shabnam and Sabetzadeh, Mehrdad and Amyot, Daniel and Liao, Jain},
  booktitle={2024 IEEE 32nd International Requirements Engineering Conference (RE)}, 
  title={Rethinking Legal Compliance Automation: Opportunities with Large Language Models}, 
  year={2024},
  volume={},
  number={},
  pages={432-440},
  doi={10.1109/RE59067.2024.00051}}

@incollection{chithaluru2020organization,
 title={Organization security policies and their after effects},
 author={Chithaluru, Premkumar and Prakash, Ravi},
 booktitle={Information security and optimization},
 pages={43--60},
 year={2020},
 publisher={Chapman and Hall/CRC}
}

@article{ulnicane2021framing,
  title={Framing governance for a contested emerging technology: insights from AI policy},
  author={Ulnicane, Inga and Knight, William and Leach, Tonii and Stahl, Bernd Carsten and Wanjiku, Winter-Gladys},
  journal={Policy and Society},
  volume={40},
  number={2},
  pages={158--177},
  year={2021},
  publisher={Oxford University Press}
}

@article{mueller2025s,
  title={It's just distributed computing: Rethinking AI governance},
  author={Mueller, Milton L},
  journal={Telecommunications Policy},
  volume={49},
  number={3},
  pages={102917},
  year={2025},
  publisher={Elsevier}
}

@article{cave2026regulating,
  title={Regulating increasingly digitalised networks and services},
  author={Cave, Martin},
  journal={Telecommunications Policy},
  volume={50},
  number={5},
  pages={103207},
  year={2026},
  publisher={Elsevier}
}

@article{micklitz2024compliance,
  title={Compliance and enforcement in the AIA through AI},
  author={Micklitz, Hans-W and Sartor, Giovanni},
  journal={Yearbook of European Law},
  volume={43},
  pages={297--341},
  year={2024},
  publisher={Oxford University Press}
}

@article{jarrahi2018humanai,
  title={Artificial intelligence and the future of work: Human-AI symbiosis in organizational decision making},
  author={Jarrahi, Mohammad Hossein},
  journal={Business horizons},
  volume={61},
  number={4},
  pages={577--586},
  year={2018},
  publisher={Elsevier}
}

@article{shrestha2019organizational,
author = {Yash Raj Shrestha and Shiko M. Ben-Menahem and Georg von Krogh},
title ={Organizational Decision-Making Structures in the Age of Artificial Intelligence},
journal = {California Management Review},
volume = {61},
number = {4},
pages = {66-83},
year = {2019},
doi = {10.1177/0008125619862257},
URL = {https://doi.org/10.1177/0008125619862257}
}

@article{coussement2024explainable,
title = {Explainable AI for enhanced decision-making},
journal = {Decision Support Systems},
volume = {184},
pages = {114276},
year = {2024},
issn = {0167-9236},
author = {Kristof Coussement and Mohammad Zoynul Abedin and Mathias Kraus and Sebastián Maldonado and Kazim Topuz}
}

@article{wang2024alert,
  title={To alert or alleviate? A natural experiment on the effect of anti-phishing laws on corporate IT and security investments},
  author={Wang, Xiaoxiao and Li, Wilson Weixun and Leung, Alvin Chung Man and Yue, Wei Thoo},
  journal={Decision Support Systems},
  volume={179},
  pages={114173},
  year={2024},
  publisher={Elsevier}
}

@article{sadeghi2024explainable,
  title={Explainable artificial intelligence and agile decision-making in supply chain cyber resilience},
  author={Sadeghi, Kiarash and Ojha, Divesh and Kaur, Puneet and Mahto, Raj V and Dhir, Amandeep},
  journal={Decision Support Systems},
  volume={180},
  pages={114194},
  year={2024},
  publisher={Elsevier}
}

\appendix
\section{Coverage Annotation Rubric}
\label{appsec:anno_rubric}

This appendix formalizes the rubric used for expert ground truth construction. 
Annotations were performed by domain experts with prior experience in 
cybersecurity governance and policy auditing. Each control was evaluated 
based on the presence, specificity, and enforceability of aligned policy evidence.

\begin{enumerate}

    \item \textbf{Fully Covered}
    \begin{itemize}
        \item Explicit articulation of control objective
        \item Clearly defined scope and applicability
        \item Assigned responsibility or ownership
        \item Defined enforcement, oversight, or review mechanism
        \item Procedural or operational guidance supporting implementation
    \end{itemize}
    
    \item \textbf{Partially Covered}
    \begin{itemize}
        \item Implicit or incomplete articulation of control intent
        \item Missing or ambiguously defined responsibility assignment
        \item Lacking enforcement or review specification
        \item Insufficient procedural clarity
        \item Coverage limited to high-level policy statements
    \end{itemize}
    
    \item \textbf{Not Covered}
    \begin{itemize}
        \item No substantive policy text aligned with control intent
        \item References present but unrelated to control requirements
    \end{itemize}

\end{enumerate}

\noindent
In cases of ambiguity, annotators adopted a conservative interpretation, 
defaulting to \textit{Partially Covered} unless explicit and enforceable 
policy evidence was present. Disagreements were resolved through 
discussion and adjudication to ensure labeling consistency.

\end{document}